\documentclass[preprint,12pt]{elsarticle}

\usepackage[T1]{fontenc}
\usepackage[english]{babel}
\usepackage{csquotes}
\usepackage{amsmath,amssymb}
\usepackage{mathtools}
\usepackage{tensor}
\usepackage{scalefnt}
\usepackage[centertableaux, smalltableaux]{ytableau}
\usepackage{graphicx}
\usepackage{listings}
\usepackage[tableposition=top]{caption}
\usepackage{booktabs}
\usepackage{acronym}
\usepackage{nth}
\usepackage{imakeidx}
\usepackage{array}
\usepackage{enumitem}
\usepackage[hidelinks,bookmarksnumbered=true,plainpages=false,linktocpage=true]{hyperref}
\usepackage[capitalise]{cleveref}

\biboptions{sort&compress}

\crefname{relation}{relation}{relations}
\creflabelformat{relation}{{\upshape(#2#1#3)}}
\definecolor{codegreen}{rgb}{.58,.69,.5}
\definecolor{codegray}{rgb}{.5,.5,.5}
\definecolor{codeblue}{rgb}{.35,.39,.6}
\definecolor{codered}{rgb}{.55,.3,.45}
\definecolor{backcolour}{rgb}{.95,.95,.95}
\newcommand\CODEstyle{\color{black}\ttfamily\scriptsize}
\lstdefinestyle{yamlModel}{
    backgroundcolor=\color{backcolour},
    commentstyle=\color{codegreen},
    keywordstyle=\color{codeblue},
    numberstyle=\color{codegray}\ttfamily\tiny,
    stringstyle=\color{codeblue},
    basicstyle=\CODEstyle,
    breakatwhitespace=false,
    breaklines=true,
    postbreak=\mbox{\textcolor{codegray}{$\hookrightarrow$}\space},
    captionpos=b,
    keepspaces=true,
    numbers=left,
    numbersep=5pt,
    showspaces=false,
    showstringspaces=false,
    showtabs=false,
    tabsize=2,
    comment=[l]{\#},
    keywords={name,description,conjugation,symmetries,lorentz_group,sun_groups,N,indices,u1_groups,fields,representations,tex,tex_hc,generations},
    moredelim=**[il][\color{codered}{:}\CODEstyle]{:},
}
\lstdefinestyle{yamlOperator}{
	style=yamlModel,
    keywords={version,type,generations,n_terms,n_operators,invariants,permutation_symmetries,vector,symmetry,matrix},
}
\lstdefinestyle{pythonSample}{
	language=Python,
    backgroundcolor=\color{backcolour},
    commentstyle=\color{codegreen},
    keywordstyle=\color{codeblue},
    numberstyle=\color{codegray}\ttfamily\tiny,
    stringstyle=\color{codeblue},
    basicstyle=\CODEstyle,
    breakatwhitespace=false,
    breaklines=true,
    postbreak=\mbox{\textcolor{codegray}{$\hookrightarrow$}\space},
    captionpos=b,
    keepspaces=true,
    numbers=left,
    numbersep=5pt,
    showspaces=false,
    showstringspaces=false,
    showtabs=false,
    tabsize=2,
    comment=[l]{\#},
}

\makeatletter
\indexsetup{level=\relax,toclevel=section,noclearpage}%
\makeindex[name=cp,title=]%
\newcommand\Texinfocommandstyletextvar[1]{{\normalfont{}\textsl{#1}}}%
\newenvironment{Texinfopreformatted}{%
  \par\GNUTobeylines\obeyspaces\frenchspacing\parskip=\z@\parindent=\z@}{}
{\catcode`\^^M=13 \gdef\GNUTobeylines{\catcode`\^^M=13 \def^^M{\null\par}}}
\newenvironment{Texinfoindented}{\begin{list}{}{}\item\relax}{\end{list}}
\setlist[description]{style=nextline, font=\normalfont}
\setlist[itemize]{label=\textbullet}
\makeatother

\newcounter{bla}

\newpageafter{abstract}

\newcommand*{\abbrev}[1]{{\scalefont{.9}#1}}
\newcommand*{\citere}[1]{Ref.~\cite{#1}}
\newcommand*{\citeres}[1]{Refs.~\cite{#1}}
\newcommand*{\code}[1]{\texttt{#1}}
\newcommand*{\samp}[1]{`\texttt{#1}'}
\newcommand*{\file}[1]{\texttt{#1}}
\newcommand*{\autoeft}{\code{AutoEFT}}
\newcommand*{\python}{\code{Python}}
\newcommand*{\sage}{\code{Sage\-Math}}
\newcommand*{\form}{\code{FORM}}
\newcommand*{\yaml}{\code{YAML}}

\newcommand*{\ytabvdots}{\none[\mathbin{\raisebox{.15em}{\rotatebox[origin=c]{90}{$\scriptscriptstyle{\dots}$}}}]}
\newcommand*{\heq}{\mathrel{\widehat{=}}}
\newcommand*{\Chi}{\raisebox{.25em}{$\boldsymbol{\chi}$}}
\newcommand*{\T}{\mathrm{T}}
\DeclareMathOperator{\cc}{C}

\newcounter{notecount}

\newcommand{\myacrodef}[3]{\acrodef{#2}{#3}\newcommand{#1}{\ac{#2}}}
\myacrodef{\sm}{SM}{Standard Model}
\myacrodef{\qft}{QFT}{Quantum Field Theory}
\acrodefplural{QFT}{Quantum Field Theories}
\newcommand{\qfts}{\acp{QFT}}
\myacrodef{\QED}{QED}{Quantum Electrodynamics}
\myacrodef{\qcd}{QCD}{Quantum Chromodynamics}
\myacrodef{\eft}{EFT}{Effective Field Theory}
\acrodefplural{EFT}{Effective Field Theories}
\newcommand{\efts}{\acp{EFT}}
\myacrodef{\smeft}{SMEFT}{Standard Model Effective Field Theory}
\myacrodef{\grsmeft}{GRSMEFT}{gravity-extension of \smeft}
\myacrodef{\LEFT}{LEFT}{Low-Energy Effective Field Theory}
\myacrodef{\wet}{WET}{Weak Effective Theory}
\myacrodef{\ibp}{IbP}{integration-by-parts}
\myacrodef{\eom}{EoM}{equation-of-motion}
\acrodefplural{EoM}{equations-of-motion}
\newcommand{\eoms}{\acp{EoM}}
\myacrodef{\mfv}{MFV}{Minimal Flavor Violation}
\myacrodef{\yd}{YD}{Young Diagram}
\acrodefplural{YD}[YDs]{Young Diagrams}

\myacrodef{\pypi}{PyPI}{Python Package Index}
\myacrodef{\uolea}{UOLEA}{Universal One-Loop Effective Action}

\usepackage{fancyhdr}
\newcommand{\RHheaderline}{\textsf{TTK-23-25~/~P3H-23-066~---~September~2023}}  %chktex 8
\fancypagestyle{pprintTitle}
{
  
  \fancyhead[R]{\RHheaderline}
}
\begin{document}

\begin{frontmatter}

\title{AutoEFT: Automated Operator Construction for Effective Field Theories}  %chktex 13

\author{Robert~V.~Harlander}
\author{Magnus~C.~Schaaf}

\address{Institute for Theoretical Particle Physics and Cosmology,\\ RWTH Aachen University, 52056 Aachen, Germany}

%- }}}
%- {{{ abstract, keywords:

\begin{abstract}
The program \autoeft\ is described. It allows one to generate \efts\ from a given
set of fields and symmetries. Allowed fields include scalars, spinors, gauge
bosons, and gravitons. The symmetries can be local or global Lie groups based
on $U(1)$ and $SU(N)$. The mass dimension of the \eft\ is limited only by the
available computing resources. The operators are stored in a compact, human
and machine-readable format. Aside from the program itself, we provide input
files for \efts\ based on the Standard Model and a number of its
extensions. These include additional particles and symmetries, \efts\ with
minimal flavor violation, and gravitons.
\end{abstract}

\begin{keyword}
EFT \sep SMEFT \sep Operator Basis%chktex 1
\end{keyword}

\end{frontmatter}

%- }}}
%- {{{ program summary, toc:

\acresetall%

{\bf PROGRAM SUMMARY}

\begin{small}
\noindent
{\em Program title\/:}
\autoeft{}

{\em Developer's repository link\/:}
\url{https://gitlab.com/auto_eft/autoeft}

{\em Licensing provisions\/:}
MIT license (MIT)

{\em Programming language\/:}
\python{}

{\em Supplementary material\/:}
\file{README.md}, \file{models.tar.xz}

{\em Nature of problem\/:} The \emph{bottom-up} construction of an \eft\ which
describes physics below a certain energy scale $\Lambda$ requires obtaining a
set of operators, composed of fields with mass $m\ll \Lambda$, that are
invariant under certain symmetries. One is primarily interested in complete
sets of independent operators, called \emph{operator bases}.  Their
construction for a given mass dimension of the operators is nontrivial due to
algebraic and kinematic relations that may render different operators
redundant.  Except for the lowest mass dimensions, the number of operators is
so large that the task of constructing an explicit \eft\ operator basis
requires a high degree of automation on a computer.  In addition, an automated
approach will allow one to immediately take into account newly postulated or
discovered light particles beyond the Standard Model.

{\em Solution method\/:} Based on the group theoretical techniques and
concepts established in \citeres{summary1,summary2,summary3,summary4,summary5,summary6,summary7,summary8,summary9} and in particular \citeres{summary10,summary11}, we developed the
program \autoeft, capable of constructing a non-redundant on-shell operator
basis for general \efts\ and arbitrary mass dimension.  Provided a suitable
\emph{model file}, the respective operator basis is generated explicitly,
including contractions of the symmetry group indices, in a fully automated
fashion.  Due to the generality of the algorithm, it can be applied to a
variety of low-energy scenarios.  The underlying low-energy theory is encoded
in a model file which defines the symmetries and the field content.  The
fairly simple format enables the user to compose their own model files and to
construct the respective operator basis with minimal effort.

{\em Additional comments including restrictions and unusual features\/:} In
its current form, \autoeft\ is restricted to theories including particles with
spin 0, 1/2, 1, and 2, where the latter two are considered massless.  In
addition, \autoeft\ only constructs operators that mediate proper
interactions, meaning that any operator must be composed of at least three
fields.  The internal symmetries must be given as factors of $U(1)$ and
$SU(N)$ groups.  In principle, operator bases can be generated for any mass
dimension, which is, however, limited by the available computing resources.

\end{small}

\acresetall%

\clearpage
\tableofcontents
\clearpage

%- }}}
%- {{{ main-text:

\begin{center}
\includegraphics[width=.3\textwidth]{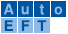}
\end{center}
\vspace{3em}

%- {{{ section{Introduction}

\section{Introduction}\label{sec:introduction}

The mass of the Higgs boson is in a region where the \sm\ remains free of
theoretical inconsistencies up to very large mass
scales~\cite{ATLAS:2022vkf,CMS:2022dwd,Hambye:1996wb,Degrassi:2012ry}.
Disregarding arguments of naturalness which have lost much of their persuasive
power due to the absence of any new particle discoveries in the TeV range, one
may have to face the possibility that on-shell discoveries of particles belong
to the past, and fundamental physics beyond the Standard Model will manifest
itself at (current or future) particle colliders only through virtual
effects~\cite{Harlander:2023kak}.

Fortunately, such effects can be parameterized in a systematic way in terms of
\efts. Ideally, the free parameters of an \eft, the Wilson coefficients, or
characteristic subsets thereof, can be determined experimentally via precision
measurements. Comparison to theoretical calculations of these coefficients via
matching to theoretical models of the heavy physics could lead to new
fundamental insights about the nature of \abbrev{UV} physics.\footnote{Dubbed
the \enquote{Cinderella approach} in \citere{Harlander:2023kak}.}

An \eft\ is based on the field content of a renormalizable Lagrangian
$\mathcal{L}_{\leq 4}$, and incorporates effects up to order ${(E/\Lambda)}^N$
in processes at energies $E$, where $\Lambda$ is the scale of new physics. The
corresponding effective Lagrangian can be written as
\begin{equation}\label{eq:preliminaries:jody}
  \begin{aligned}
    \mathcal{L} &= \mathcal{L}_{\leq 4}
    + \sum_{d=5}^{N+4}\sum_n\frac{\mathcal{C}^{(d)}_{n}}{\Lambda^{d-4}}\mathcal{O}^{(d)}_n\,,
  \end{aligned}
\end{equation}
where the higher-dimensional operators $\mathcal{O}^{(d)}_n$ are composed of
all fields of $\mathcal{L}_{\leq 4}$, and $\mathcal{C}^{(d)}_n$ are the Wilson
coefficients. For example, if $\mathcal{L}_{\leq 4}$ is the \sm\ Lagrangian,
then the higher-dimensional operators are composed of all \sm\ fields, and
$\mathcal{L}$ is referred to as the \smeft\ Lagrangian.

In a top-down approach, the effective operators follow from a
\abbrev{UV}-complete theory $\mathcal{L}_\mathrm{UV}$ by integrating out the
heavy degrees of freedom in the path integral of the generating
functional. Such an approach is pursued in the \uolea, for example, which
also provides the Wilson coefficients in terms of the parameters of
$\mathcal{L}_\mathrm{UV}$~\cite{Drozd:2015rsp,Ellis:2017jns,Kramer:2019fwz,Banerjee:2023iiv,Chakrabortty:2023yke}.

More common, however, is a bottom-up approach, which will be adopted in
this paper. Here, one constructs all higher-dimensional operators by combining
the fields of the low-energy theory $\mathcal{L}_{\leq 4}$ in such a way that
they obey all symmetry constraints. At the same time, however, one requires
the set of operators to be non-redundant in order to ensure that the Wilson
coefficients are well-defined. Redundancies among the set of operators can
arise from several sources. First, since total derivatives in the Lagrangian
do not contribute to the action, operators could be related by
\ibp\ identities. Second, operators could be linearly dependent due to
algebraic relations such as Fierz or Schouten identities. Third,
higher-dimensional operators that vanish due to \eoms\ can be eliminated from
the \eft\ by field redefinitions in the path integral of the generating
functional~\cite{Arzt:1993gz,Criado:2018sdb}.
Finally, operators could be related by permutations of the fields transforming
in equal representations~\cite{Fonseca:2019yya}.

Obviously, the \eft\ which is most relevant from a phenomenological point of
view is \smeft.  In fact, the dimension-five operators are strong candidates
for being the source of neutrino masses~\cite{Weinberg:1979sa}. The multiple attempts needed to
arrive at the \smeft-bases at dimension six and seven testify to the
complexity of constructing a complete and non-redundant set of operators,
despite the fact that their numbers are still quite manageable (84 and 30,
respectively, for a single generation of
fermions)~\cite{Buchmuller:1985jz,Grzadkowski:2010es,
  Lehman:2014jma,Liao:2016hru}.\footnote{See
\cref{ssec:counting-smeft-operators} concerning the counting of operators.}
Towards higher mass dimension, this number increases roughly exponentially. It
can actually be computed exactly using Hilbert-series
techniques~\cite{Lehman:2015via,Lehman:2015coa,Henning:2015daa,
  Henning:2015alf,Marinissen:2020jmb,Banerjee:2020bym,Calo:2022jqv}, but also
by more direct
methods~\cite{Gripaios:2018zrz,Fonseca:2017lem,Criado:2019ugp}.\footnote{See
also \citere{Aebischer:2023irs} for a summary of \eft\ software tools.}
Despite the fact that, for three generations of fermions, the number of
operators in the \smeft\ basis at mass dimensions eight and nine already
amounts to 44807 and 90456, respectively, it was still possible to construct
them by largely manual
efforts~\cite{Murphy:2020rsh,Liao:2020jmn}. Nevertheless, an algorithmic
procedure clearly becomes desirable.

Important steps towards the systematic construction of \eft\ operators were
made in \citeres{Shadmi:2018xan,Henning:2019mcv,Ma:2019gtx,Henning:2019enq,
  Aoude:2019tzn,Fonseca:2019yya,Durieux:2019eor,Falkowski:2019zdo,
  Durieux:2019siw,Fonseca:2020vke}. A complete algorithm was presented in
\citeres{Li:2020gnx,Li:2020xlh} and used to construct the \smeft\ basis at
mass dimensions eight and nine. Its implementation is available as a
\code{Mathematica} package~\cite{Li:2022tec}.
In \citere{Harlander:2023psl}, we reported on an
independent implementation of that algorithm and used it to derive for the
first time the \smeft\ operator bases at mass dimensions 10, 11, and 12. The
current paper accompanies the publication of the associated computer program,
named
\autoeft.\footnote{\label{foot:repo}\url{https://gitlab.com/auto_eft/autoeft}}
It is available as open source under the
MIT~license,\footnote{\url{https://spdx.org/licenses/MIT.html}} is based on
\python,\footnote{\url{https://www.python.org/}} and uses only publicly
available software libraries, in particular
\sage.\footnote{\url{https://www.sagemath.org}} Since the algorithm of
\citeres{Li:2020xlh,Li:2020gnx} is not specific to \smeft, it is possible to
use \autoeft\ also in extended theories with additional light particles beyond
the \sm\ spectrum (see, e.g.,
\citeres{Liao:2016qyd,Liao:2020zyx,Li:2020tsi,Li:2021tsq,
  Li:2023wdz,Song:2023lxf,Song:2023jqm,Liang:2023yta}). For example,
\citere{Harlander:2023psl} also includes the operators of the \grsmeft~\cite{Ruhdorfer:2019qmk} up to
mass dimension 12.

This paper provides an introduction to \autoeft, describing the necessary
notation, the preparation of the input file, the commands to generate the
operator basis, and the format of the output files.  \Cref{sec:preliminaries}
introduces the theoretical and notational background required to interpret the
\autoeft\ input and output.  The installation of \autoeft\ is described in
\cref{sec:installation}.  \Cref{sec:the-model-file} explains the structure of
the model file to be processed by \autoeft, and provides comprehensive
examples for various models.  The operator construction using \autoeft\ is
showcased in \cref{sec:constructing-operators}, including a discussion on the
output format, as well as \autoeft's current limitations.
\Cref{sec:operator-files} contains examples on how \autoeft\ can further
process the output. In addition, we include a reference manual in the
appendix, which can be used to look up particular features or specifications
related to the usage of \autoeft.

%- }}}
%- {{{ section{Preliminaries}

\section{Preliminaries}\label{sec:preliminaries}

For fixed values of the Wilson coefficients and the parameters of the
low-energy theory, an \eft\ can be considered as a vector in the space of all
higher-dimensional operators\@. \autoeft\ constructs a basis in this space of
operators for a fixed, but in principle arbitrary value of $d$. In doing so,
it takes into account the constraints arising from external (Lorentz) and
internal symmetries. It ensures that the basis is non-redundant, meaning that
no two operators are interrelated through \eoms, \ibp\ or algebraic
identities.

The field content and the symmetry groups of the low-energy theory
$\mathcal{L}_{\leq 4}$ are supplied to \autoeft\ via an input file, referred
to as \emph{model file} in the following. Its detailed structure will be
defined in \cref{sec:the-model-file} and \labelcref{anchor:Model}. In this
section, we provide the notational background for its contents.

\autoeft\ allows for particles with spin 0, 1/2, 1, and 2
in the spectrum of $\mathcal{L}_{\leq 4}$.\footnote{We use \enquote{fields} and
\enquote{particles} interchangeably in this paper.} For spin~0 and spin~1/2, it makes
no difference for the construction of an \eft~\cite{Li:2020tsi} whether they
are massive or massless, and thus also for \autoeft. Higher-spin particles
represented by vector or tensor fields are currently restricted to the
massless case though. Note that this is in line with \smeft, which is
formulated in the unbroken phase of the \sm\ Lagrangian. The massive vector
bosons are recovered by performing the electroweak symmetry breaking in
\smeft\ explicitly. Furthermore, since scalar and spinor fields are allowed to
be massive, \autoeft\ can also be used to generate \efts\ in which all massive
vector bosons are integrated out (e.g., \LEFT/\wet, parameterizing effects
between the electroweak scale and $\Lambda_{\text{QCD}}$).  For \autoeft, a
particle is thus uniquely identified by its $U(1)$ charges, the
representations according to which it transforms under the Lorentz and the
non-abelian internal symmetry groups, and a possible generation index.

The irreducible representations of the Lorentz group---which can be identified
with ${SU(2)}_l\times{SU(2)}_r$ for our purpose---are characterized by
$(j_l,j_r)$, where $j_{l/r}$ are non-negative integers or half-integers.  The
most important irreducible representations are given by $(0,0)$, $(1/2,0)$,
and $(1,0)$, corresponding to scalars $\phi$, left-handed Weyl spinors
${\psi_{\mathrm{L}}}_\alpha$, and self-dual 2-forms
$F_{\mathrm{L}\,\alpha\beta}$. For simplicity, we will refer to the latter
also as \enquote{left-handed field-strength tensors} in the following.  In
addition, we consider self-dual (\enquote{left-handed}) Weyl tensors
$C_{\mathrm{L}\,\alpha\beta\gamma\delta}$ transforming as $(2,0)$ which are
required for gravity. Since $j_r=0$ for all of these \enquote{elemental}
representations, they can also be characterized by their \emph{helicity}
\begin{equation}\label{eq:helicity}
	h=j_r - j_l\,.
\end{equation}
The conjugate (\enquote{right-handed}) fields ${\psi_\mathrm{R}}^{\dot{\alpha}}$,
$F_\mathrm{R}^{\,\dot{\alpha}\dot{\beta}}$, and
$C_\mathrm{R}^{\,\dot{\alpha}\dot{\beta}\dot{\gamma}\dot{\delta}}$ transform as
$(0,1/2)$, $(0,1)$ and $(0,2)$ under the Lorentz group and thus carry the
negative helicity of the corresponding left-handed fields.  Here and in
the following, $\alpha$, $\beta$, \ldots\ and $\dot{\alpha}$, $\dot{\beta}$,
\ldots\ denote fundamental ${SU(2)}_l$ and ${SU(2)}_r$ spinor indices,
respectively, unless indicated otherwise.\footnote{We do not consider the
irreducible representation $(1/2,1/2)$ corresponding to Lorentz four-vectors
explicitly, because we assume that vector fields always arise as gauge fields
and thus appear only as part of a field-strength tensor or the covariant
derivative. For more details, see the subsequent main text.}

All other fields which occur in common \qfts\ transform in representations
which can be composed of these elemental representations $(|h|,0)$ and their
conjugate versions $(0,|h|)$. For example, the bispinor and the field-strength
tensor transform in the direct sums of the left- and right-handed Weyl spinor
representations $(1/2,0)\oplus(0,1/2)$, and the self- and anti-self-dual
2-form representations $(1,0)\oplus(0,1)$, respectively.  In \autoeft,
however, one simply defines each irreducible component as a separate field.
Concrete examples will be given in \cref{sec:the-model-file}.

\begin{table}
\centering
\caption{ The list of irreducible representations of the Lorentz group
  supported by \autoeft.  Each representation is associated with a placeholder
  symbol for the field, and a unique value for the helicity $h$.}%
\label{tab:helicities}
\begin{tabular}{cccc}
\toprule
field & $(j_l,j_r)$ & $h$ & name \\
\midrule
$\phi$ & $(0,0)$ & $0$ & scalar \\
$\psi_{\mathrm{L}}$ & $(1/2,0)$ & $-1/2$ & left-handed spinor \\
$\psi_{\mathrm{R}}$ & $(0,1/2)$ & $+1/2$ & right-handed spinor \\
$F_{\mathrm{L}}$ & $(1,0)$ & $-1$ & left-handed field-strength tensor \\
$F_{\mathrm{R}}$ & $(0,1)$ & $+1$ & right-handed field-strength tensor \\
$C_{\mathrm{L}}$ & $(2,0)$ & $-2$ & left-handed Weyl tensor \\
$C_{\mathrm{R}}$ & $(0,2)$ & $+2$ & right-handed Weyl tensor \\
\bottomrule
\end{tabular}
\end{table}

The output of \autoeft\ is thus formulated in terms of the objects summarized
in \cref{tab:helicities}, as well as the covariant derivative
$D^{\dot{\alpha}}_\alpha$. The action of $n$ derivatives on a field $\Phi$ is
understood in the \autoeft\ output as the combined object
\begin{equation}\label{eq:covD}
	{(D^n \Phi)}_{(\alpha\beta\dots)}^{(\dot\alpha\dot\beta\dots)}
	\sim {(D^n \Phi)}_{\alpha\beta\dots}^{\dot\alpha\dot\beta\dots}
	+ {(D^n \Phi)}_{\beta\alpha\dots}^{\dot\alpha\dot\beta\dots}
	+ {(D^n \Phi)}_{\alpha\beta\dots}^{\dot\beta\dot\alpha\dots}
	+ {(D^n \Phi)}_{\beta\alpha\dots}^{\dot\beta\dot\alpha\dots}
	+ \dots\,,
\end{equation}
where the dotted and undotted indices are separately symmetrized.  For each
term on the right hand side of \cref{eq:covD}, the first $n$ pairs of dotted
and undotted indices belong to the covariant derivatives, whereas all
remaining indices are part of the field $\Phi$.

In order to facilitate the translation of the operators into the more common
notation of bispinors $\Psi$, field-strength tensors $F^{\mu\nu}$, Weyl
tensors $C^{\mu\nu\rho\sigma}$, and covariant derivatives $D^\mu$, with
Lorentz four-vector indices $\mu,\nu,\dots$, we collect the necessary
relations in \labelcref{sec:relations-to-conventional-notation}.

Concerning internal symmetries, \autoeft\ allows for local and global $U(1)$
and $SU(N)$ groups.\footnote{Concerning $U(N)$, see \cref{ssec:mfv-model}.}
All fields are assumed to transform in an irreducible representation of the
internal symmetry groups\@. \autoeft\ requires that each $U(1)$ charge of a
field is given by a (fractional) multiple of some elementary charge (which
does not need to be specified further).  In the model file, the $U(1)$ charges
are thus defined by rational numbers.  Examples will be given in
\cref{sec:the-model-file}.

The irreducible representations of $SU(N)$ are encoded via their one-to-one
correspondence to Young diagrams, which can be represented by
lists of non-increasing positive integers (also referred to as \emph{integer
partitions} in the following).\footnote{\label{foot:dynkin1} \autoeft\ also supports the characterization of these representations by Dynkin labels; see \cref{sssec:qcd}, for a concrete example.}
For example, the fundamental representation
of $SU(N)$ can be specified as
\begin{equation}\label{eq:sun:fundamental_repr}
	\ydiagram{1} \quad\sim\quad [1] \,.
\end{equation}
For the anti-fundamental representation, it is
\begin{equation}\label{eq:sun:anti_fundamental_repr}
    \rotatebox[origin=c]{90}{$\scriptstyle{N-1}$}
    \left\{\vphantom{\ydiagram{1,1,1,1}}\right.
    \begin{ytableau}
        ~ \\
        ~ \\
        \ytabvdots \\
        ~
    \end{ytableau}
	\quad\sim\quad
	    [\underbrace{1,1,\dots,1}_{N-1}]
	    \equiv [1^{N-1}]\,,
\end{equation}
and for the adjoint representation, the correspondence is
\begin{equation}\label{eq:sun:adjoint_repr}
    \rotatebox[origin=c]{90}{$\scriptstyle{N-1}$}
    \left\{\vphantom{\ydiagram{1,1,1,1}}\right.
    \begin{ytableau}
        ~ & ~ \\
        ~ \\
        \ytabvdots \\
        ~
    \end{ytableau}
	\quad\sim\quad
	    [2,\underbrace{1,\dots,1}_{N-2}]
            \equiv [2,1^{N-2}]\,,
\end{equation}
where we used a common short-hand notation for integer partitions with long
sequences of the same number.

In the \sm\ context, it is more common to refer to the irreducible
representations of $SU(N)$ by their dimensionality rather than by integer
partitions or Dynkin labels. For example, $\mathbf{3}$, $\mathbf{\bar{3}}$, and
$\mathbf{8}$ denote the fundamental, anti-fundamental, and adjoint representations
of $SU(3)$, respectively. However, this characterization becomes ambiguous in the
case where several non-equivalent irreducible representations with the same
dimensionality exist (e.g., $SU(3)$ has four 15-dimensional irreducible
representations: $[4], [4,4], [3,1], [3,2]$). Such a characterization is,
therefore, not suitable for a systematic approach, and we refrain from it in
the context of \autoeft.

Similar to the Lorentz group, the fields composing the operators in the output
of \autoeft\ carry only fundamental indices of the internal symmetry groups.
For fields transforming in the anti-fundamental or adjoint representations,
one can translate this directly to a more common notation using the relations
provided in \labelcref{sec:relations-to-conventional-notation}. While this is
sufficient for \smeft, it may be desirable to translate other representations
in extended theories with light fields. In this case, the corresponding
Clebsch-Gordan coefficients need to be taken into account.\footnote{For
example, the sextet representation $\ydiagram{2} \sim [2]$ of $SU(3)$ can be
related to the symmetric product of two fundamental representations using the
Clebsch-Gordan coefficients computed in \citere{Han:2009ya}.  Consequently, a
field transforming in this representation can be denoted either by one sextet
index or two fundamental indices, related by the Clebsch-Gordan coefficients.}

%- }}}
%- {{{ section{Installation}

\section{Installation}\label{sec:installation}

%- {{{ intro:

\autoeft\ is implemented in \python\ and makes use of several
functions provided by the free open-source mathematics software system
\sage. Since intermediate expressions during the construction
procedure can become exceedingly large, certain algebraic operations are
passed to \form~\cite{Vermaseren:2000nd,Kuipers:2012rf}.  All remaining
dependencies are third-party \python\ libraries and are included for the
user's convenience, such as input validation and console markup.  For a
standard installation of \autoeft, the following software needs to be
installed on the system:
\begin{description}[style=nextline]
\item[\python\ (\textit{version 3.8} or later)]
This requirement is fulfilled by default in most cases.
There is either a system wide \python\ installation that is also used by \sage, or \sage\ does come with its own version of the \python\ interpreter.
If the installation is done via the \code{conda}/\code{mamba} package management system, a suitable \python\ version is automatically included in the virtual environment.

\item[\sage\ (\textit{version 9.3} or later)] The \sage\ library only needs to
  be installed explicitly if \autoeft\ is \emph{not} installed using the
  \code{conda}/\code{mamba} package management system.\footnote{Although
  there is some effort towards modularizing \sage\ into separate
  distributions, the packages required by \autoeft\ are only available in the
  complete library for now.  We advice to either install \sage\ by
  \enquote{hand}, or to use the \code{conda}/\code{mamba} package management
  system, which installs \sage\ automatically in a virtual environment.  }
  Installation details can be found at
  \url{https://doc.sagemath.org/html/en/installation/index.html}.

\item[\form\ (\textit{version 4.3} or later)]
The \form\ home page can be found at \url{https://www.nikhef.nl/~form/}.
To use \autoeft\ together with \form, make sure that there is an executable named \code{form} on the system path or on a path specified by the environment variable \code{AUTOEFT\_PATH} (cf.~\labelcref{anchor:Environment}).
\end{description}

In case of problems with the installation, the user is advised to contact the
authors via email or the \autoeft~repository, see \cref{foot:repo}. The latter also
collects several potential installation issues and their resolution.

%- }}}
%- {{{ subsection{Installing AutoEFT from PyPI}

\subsection{Installing \autoeft\ from \texttt{PyPI}}\label{ssec:install-from-pypi}
This is the recommended installation method. It requires an existing and
running version of \sage\ though.  Given that, \autoeft\ and its dependencies
can be installed from the \textit{\pypi}\footnote{\url{https://pypi.org/}} by
simply running:%
\footnote{On \code{macOS} using \code{Homebrew}, it may be
necessary to precede this statement by
\code{PYTHONEXECUTABLE=</path/to/sage>}
with the proper path to the \sage\ executable inserted.
In addition, it may be necessary to add the path to
\sage's executables to the \code{\$PATH} environment variable.}
\begin{Texinfoindented}
\begin{Texinfopreformatted}%
\ttfamily sage -pip install autoeft
\end{Texinfopreformatted}
\end{Texinfoindented}

%- }}}
%- {{{ subsection{Installing AutoEFT from conda-forge}

\subsection{Installing \autoeft\ from \texttt{conda-forge}}\label{ssec:install-from-conda-forge}
Since the \sage\ distribution is part of the \textit{conda-forge}~\cite{conda_forge_community_2015_4774216} channel, there is no requirement for a prior installation.
Using the \code{conda}\footnote{\url{https://conda.io/}} package manager, \autoeft\ and its dependencies can be installed from the \textit{conda-forge} channel by running:
\begin{Texinfoindented}
\begin{Texinfopreformatted}%
\ttfamily conda install autoeft -c conda-forge
\end{Texinfopreformatted}
\end{Texinfoindented}
If the \code{mamba}\footnote{\url{https://github.com/mamba-org/mamba}} package manager is used instead, the \textit{conda-forge} channel is enabled by default.
Hence, \autoeft\ and its dependencies can be installed by running:
\begin{Texinfoindented}
\begin{Texinfopreformatted}%
\ttfamily mamba install autoeft
\end{Texinfopreformatted}
\end{Texinfoindented}

%- }}}
%- {{{ subsection{Building AutoEFT from the Source Code}

\subsection{Building \autoeft\ from Source Code}\label{ssec:build-from-source}
To build \autoeft\ from its source code, make sure the latest version of the
Python Packaging Authority's
\code{build}\footnote{\url{https://pypi.org/project/build/}} is installed.
The distribution packages can then be generated by running:
\begin{Texinfoindented}
\begin{Texinfopreformatted}%
\ttfamily git clone https://gitlab.com/auto\_eft/autoeft.git autoeft
\ttfamily cd autoeft/
\ttfamily python -m build
\end{Texinfopreformatted}
\end{Texinfoindented}
Note that the last command must be executed in the directory containing the
file \file{pyproject.toml}.  After this, there should be two archive files in
the newly created \file{dist/} directory: The source distribution
\file{autoeft-1.0.0.tar.gz} as well as the build distribution
\file{autoeft-1.0.0-py3-none-any.whl}.  To install the local package, run:
\begin{Texinfoindented}
\begin{Texinfopreformatted}%
\ttfamily sage -pip install dist/autoeft-1.0.0-py3-none-any.whl
\end{Texinfopreformatted}
\end{Texinfoindented}
As \autoeft\ is developing, the version number will have to be replaced
accordingly in these commands, of course.

%- }}}
%- {{{ subsection{Validating the Installation}

\subsection{Validating the Installation}
A successful installation of \autoeft\ can be validated by running
\begin{Texinfoindented}
\begin{Texinfopreformatted}%
\ttfamily autoeft check
\end{Texinfopreformatted}
\end{Texinfoindented}
In the current version, this constructs the \smeft\ operator basis for mass
dimension six and compares it to a pre-constructed result.

%- }}}
%- }}}
%- {{{ section{Model File}

\section{The Model File}\label{sec:the-model-file}

%- {{{ intro:

To construct an \eft\ operator basis, the user must define a model describing
the relevant details of the low-energy theory.  This is done via the
\emph{model file} which encodes all information about the symmetries and field
content of the model.\footnote{Technically, the format of the model file is
\yaml\ (\url{https://yaml.org/}); all required specifications will be
implicitly discussed below though.}  A detailed description of all keywords
and their type can be found in \labelcref{anchor:Model}.

%- }}}
%- {{{ subsection{Basic Structure}

\subsection{Basic Structure}\label{ssec:description}
A valid model file has to contain a minimal set of keywords (simply referred
to as \emph{keys} in the following), which must be assigned appropriate
values. In particular, every model file must contain the key \code{name}, set
to a valid string that identifies the model.  The other required keys are
\code{symmetries} and \code{fields}.  These three keys are sufficient to
define a valid model file that \autoeft\ can process.  For example, in
\cref{lst:min}, both \code{symmetries} and \code{fields} are set to the empty
set \samp{\{\}}, corresponding to the trivial model without any
fields.\footnote{We adopt the convention that variable input provided by the
user is set in type-writer font and surrounded by single quotes in the main
text. In the code listings, they are set in black color. The single quotes are
missing for fixed code words that are not to be changed by the user (blue
color in the listings).}
\begin{lstlisting}[caption={Minimal data required in a
    model file.},label={lst:min},style=yamlModel]
# AutoEFT model file
name: Minimal Model
symmetries: {}
fields: {}
\end{lstlisting}

As a non-trivial model, let us consider scalar \QED, i.e.\ a $U(1)$ gauge
theory of a charged scalar field.  The $U(1)$ symmetry is implied by adding
the sub-key \code{u1\_groups} to \code{symmetries}, as displayed in \cref{lst:sqed:sym}.
\begin{lstlisting}[linerange={3-5},firstnumber=3,caption={Symmetry definition of  %chktex 8
    the scalar \QED\ model
    file.},label={lst:sqed:sym},style=yamlModel]
symmetries:
  u1_groups:
    QED: {}
\end{lstlisting}%chktex 8

Note that the actual symmetry, identified by the string \samp{QED}, has been
added as another sub-key to \code{u1\_groups}. In principle, we could specify
additional attributes for this group (e.g., an allowed violation, a residual charge, or a \LaTeX\ symbol; see
\labelcref{anchor:Model}) by assigning it a non-trivial value. For our
purposes, however, this is not necessary and we assign to it the empty set
\samp{\{\}}.

Next, we include a single complex scalar field $\phi$ in the model, by adding
the entry \samp{phi} to \code{fields} as shown in \cref{lst:sqed:phi}.
\begin{lstlisting}[linerange={6-9},firstnumber=6,caption={Definition of the
scalar field in the model
file.},label={lst:sqed:phi},style=yamlModel]
fields:
  phi:
    representations:
      QED: -1
\end{lstlisting}%chktex 8
Again, the name \samp{phi} is arbitrary. To define the transformation
properties of the field under the symmetry groups, the key
\code{representations} must be added to \samp{phi}. In our case, there is only
one symmetry group, so we add the entry \samp{QED:~-1} to
\code{representations}, which means that $\phi$ carries one negative unit of
the elementary $U(1)$ charge, see line~9 in \cref{lst:sqed:phi}\@. For every
field defined in the model file, \autoeft\ automatically takes into account
the conjugate version and denotes it by appending the symbol \samp{+} to the
original field name. Thus, in our example, the conjugate field $\phi^\dagger$
is taken into account automatically by \autoeft, and it will be denoted by
\samp{phi+} in the output.\footnote{ The exception to this are fields all of
whose representations are real (or combine to form a real representation). In
this case, no conjugate field is generated. One can also prevent
\autoeft\ from including the conjugate field---for whatever reason one may
have---by using the \code{conjugate} property; see \labelcref{anchor:Model}.
}

To make this theory an actual gauge theory, the $U(1)$ gauge boson has to be
defined as well.  Gauge bosons can appear in two instances: encoded in field
strength tensors or as part of the covariant derivative.  The latter is
automatically included by \autoeft, while the former is decomposed into two
separate fields which transform in irreducible representations of the
Lorentz group, see \cref{sec:preliminaries}. The first one,
\samp{FL}\,$\heq\,F_{\mathrm{L}}$, transforming as $(1,0)$,
can be defined in the model file by adding another entry to \code{fields}; see \cref{lst:sqed:FL}.
\begin{lstlisting}[linerange={10-12},firstnumber=10,caption={Definition of the  %chktex 8
    gauge boson in the model
    file.},label={lst:sqed:FL},style=yamlModel]
  FL:
    representations:
      Lorentz: -1
\end{lstlisting}
By default, the Lorentz group is identified by the literal string
\samp{Lorentz},%
\footnote{ This can be overwritten by the user in the model file; see
\labelcref{anchor:Model}.  } and it is assigned the helicity value $h=-1$ in
this case.  The second component of the \QED\ field-strength tensor,
$F_{\mathrm{R}} \in (0,1)$, is again included automatically, as it is the
conjugate of $F_{\mathrm{L}}$ ($F_{\mathrm{R}} = F_{\mathrm{L}}^\dagger$).
Note that we did not explicitly have to specify the helicity for the scalar
field $\phi$ in \cref{lst:sqed:phi}, nor the \QED\ charge for the field
strength $F_{\mathrm{L}}$ in \cref{lst:sqed:FL}\@. If unspecified,
\autoeft\ assumes that the fields are singlets under the corresponding
symmetry groups, which means that $\phi$ is defined as a Lorentz scalar, and
$F_\mathrm{L}$ does not carry a $U(1)$ charge, as desired.

Combining the symmetry definition in \cref{lst:sqed:sym} and the field content
definitions in \cref{lst:sqed:phi,lst:sqed:FL}---and giving the model a
suitable name---results in the entire model file, displayed in \cref{lst:sqed}.
\begin{lstlisting}[caption={Scalar \QED\ model
    file.},label={lst:sqed},style=yamlModel,float]
# AutoEFT model file
name: sQED-EFT
symmetries:
  u1_groups:
    QED: {}
fields:
  phi:
    representations:
      QED: -1
  FL:
    representations:
      Lorentz: -1
\end{lstlisting}

Note that, in the model file, an explicit association of the field strength
tensor \samp{FL} (or \samp{FR}) to the gauge group \samp{QED} is not
necessary. Its role as a gauge field will originate from the proper
interpretation of the covariant derivative in the resulting operators. If it
includes the photon field, the symmetry is local; otherwise, it is a global
symmetry, and \samp{FL} represents a vector boson which transforms as a
singlet under the symmetry group. It will still couple to the fermion in
higher-dimensional operators.

%- }}}
%- {{{ subsection{Realistic Examples}

\subsection{Realistic Examples}\label{ssec:examples}

%- {{{ intro:

In this section, more realistic examples will be considered, starting from
\QED, generalizing to \qcd, and finally the \sm. In the course of this, we
will discuss the definition of spinors and non-abelian $SU(N)$ symmetry groups
in the model file.

%- }}}
%- {{{ subsubsection{QED}

\subsubsection{QED}\label{sssec:qed}
To promote the example of scalar \QED\ from the previous section to actual
\QED, one needs to introduce Dirac fermions.  As described in
\cref{sec:preliminaries}, the Lorentz representation of bispinors is given by
$(1/2,0)\oplus(0,1/2)$.  The model file for \QED\ with a single charged
electron can thus be written as shown in \cref{lst:qed}.
\begin{lstlisting}[caption={\QED\ model
    file.},label={lst:qed},style=yamlModel,float]
# AutoEFT model file
name: QED-EFT
description: Effective Field Theory of QED interactions

symmetries:
  u1_groups:
    QED: {}

fields:
  eL: # EL = (eL, 0)^T, ELbar = (0, eL+)
    representations:
      Lorentz: -1/2
      QED: -1
  eR: # ER = (0, eR)^T, ERbar = (eR+, 0)
    representations:
      Lorentz: 1/2
      QED: -1
  FL:
    representations:
      Lorentz: -1
\end{lstlisting}
Here, $e_{\mathrm{L}}\heq$~\samp{eL} and $e_{\mathrm{R}}\heq$~\samp{eR} denote left- and right-handed Weyl spinors
$e_{\mathrm{L}}\in(1/2,0)$ and $e_{\mathrm{R}}\in(0,1/2)$ with helicity $-1/2$
and $+1/2$, respectively. In \QED, both of them carry the same charge, and
thus can be considered as components of the same Dirac spinor
\begin{equation}
  \Psi = \begin{pmatrix}
    e_\mathrm{L}\\
    e_\mathrm{R}
  \end{pmatrix}\,.
\end{equation}
\autoeft\ by default also takes into account the conjugate Weyl spinors
\samp{eL+}\,$\heq\,e_\mathrm{L}^\dagger\in (0,1/2)$ and
\samp{eR+}\,$\heq\,e_\mathrm{R}^\dagger\in (1/2,0)$. Note that for a Dirac
spinor $e_\mathrm{L}^\dagger\neq e_\mathrm{R}$, which is why both
$e_\mathrm{L}$ and $e_\mathrm{R}$ need to be defined in the model file. In
contrast, a Majorana spinor is represented in the model file by a single Weyl
spinor which transforms in real representations of all internal symmetries.

In the literature it is quite common to adopt the \emph{all-left} chirality
notation for the fundamental building blocks of an \eft.  In this convention,
all Weyl spinors are defined to be left-handed, so that the index \enquote{L} can be
dropped.  The right-handed components are then acquired by conjugation.  In
the above example this would mean that one defines
\samp{e}\,$\heq\,e\equiv e_\mathrm{L}$ and its charge conjugate
\samp{eC}\,$\heq\, e_\mathbb{C}\equiv e_{\mathrm{R}}^\dagger$. One
could thus define \QED\ in \autoeft\ also by replacing lines 10--17 in
\cref{lst:qed} by the content of \cref{lst:allleftqed}.
\begin{lstlisting}[linerange={10-17},firstnumber=10,  %chktex 8
  caption={All-left notation for the electron.},
  label={lst:allleftqed},style=yamlModel]
  e: # EL = (e, 0)^T, ELbar = (0, e+)
    representations:
      Lorentz: -1/2
      QED: -1
  eC: # ER = (0, eC+)^T, ERbar = (eC, 0)
    representations:
      Lorentz: -1/2
      QED: 1
\end{lstlisting}

%- }}}
%- {{{ subsubsection{QCD}

\subsubsection{QCD}\label{sssec:qcd}
To generalize the example of \QED\ to a non-abelian theory like \qcd, $SU(N)$
symmetries need to be introduced.  They are defined in a similar way to $U(1)$
symmetries in the model file but require additional information like their
degree $N$. A model file for \qcd\ with a single quark flavor could be defined as
displayed in \cref{lst:qcd}.
\begin{lstlisting}[caption={\qcd\ model
    file.},label={lst:qcd},style=yamlModel,float]
# AutoEFT model file
name: QCD-EFT
description: Effective Field Theory of QCD interactions

symmetries:
  sun_groups:
    QCD:
      N: 3

fields:
  qL:
    representations:
      Lorentz: -1/2
      QCD: [1]
  qR:
    representations:
      Lorentz: 1/2
      QCD: [1]
  GL:
    representations:
      Lorentz: -1
      QCD: [2,1]
\end{lstlisting} The $SU(3)$
symmetry of \qcd\ is imposed by the lines 5--8. Under the keyword \code{sun\_groups}, all
$SU(N)$ symmetry groups of the model are listed; here, we only have \samp{QCD}, for
which we specify the degree by the entry \samp{N:~3} (note the indentation
of line 8).

Lines 11--22 declare the field content of the model.  Analogously to the
example of \QED\ discussed in \cref{sssec:qed}, a Dirac quark spinor is
implemented by specifying its left- and right-handed components, named
$q_\mathrm{L}\heq$~\samp{qL} and $q_\mathrm{R}\heq$~\samp{qR} here. The fact
that they transform in the fundamental representation of \qcd\ is encoded by
specifying the integer partition \samp{[1]} in lines 14 and 18,
cf.~\cref{eq:sun:fundamental_repr}.  As discussed above,
\autoeft\ automatically takes into account the corresponding conjugate fields
\samp{qL+}\,$\heq\,q_\mathrm{L}^\dagger$ and
\samp{qR+}\,$\heq\,q_\mathrm{R}^\dagger$ which transform in the
anti-fundamental representation $[1,1]$ of \qcd,
cf.~\cref{eq:sun:anti_fundamental_repr}.  Since the adjoint representation
$[2,1]$ is real, only the left-handed component of the gluon field-strength
tensor \samp{GL} must be defined in the model file explicitly, see lines~19--22 of
\cref{lst:qcd}.

Instead of integer partitions, one may also use \emph{Dynkin labels} to
specify the irreducible representation of $SU(N)$ in which a field
transforms. For \autoeft, the difference is indicated by using round brackets
instead of square ones.  The fundamental and adjoint representations of
$SU(3)$ are denoted by the Dynkin labels $(10)$ and $(11)$, respectively.
Lines~14 and 18 of \cref{lst:qcd} could thus also be written as \samp{QCD:
  (1,0)}, for example, and line~22 as \samp{QCD: (1,1)}.\footnote{This is not
to be confused with the $(j_r,j_l)$ notation for the representations of the
Lorentz group defined in \cref{sec:preliminaries}.}  Internally, any Dynkin
label is converted to the respective partition.

%- }}}
%- {{{ subsubsection{Standard Model}

\subsubsection{Standard Model}\label{sssec:standard-model}
The previous sections provide all the information required to compose a model
file for the entire \sm\ in the unbroken phase---including all symmetries and
fields. The transition to the broken phase can be performed at the level of
the operators by appropriate replacements of the Higgs field.

The \sm\ gauge group is given by $SU(3) \times SU(2) \times U(1)$ which can be
defined in just a few lines in the model file, see lines~6--10 in
\cref{lst:sm} below.  Each gauge group is equipped with an associated
multiplet of gauge bosons by defining the components
\samp{GL}\,$\heq\,G_{\mathrm{L}}$,
\samp{WL}\,$\heq\,W_{\mathrm{L}}$, and
\samp{BL}\,$\heq\,B_{\mathrm{L}}$, respectively (cf.~lines~13--23).

The matter fields of the \sm\ come in five distinct representations. Taking
the first generation of fermions as an example, they are characterized by the
Weyl spinors
\begin{equation}\label{eq:sm:fermions}
  \text{\samp{QL}}\,\heq\,Q_{\mathrm{L}}\,,\quad
  \text{\samp{uR}}\,\heq\,u_{\mathrm{R}}\,,\quad
  \text{\samp{dR}}\,\heq\,d_{\mathrm{R}}\,,\quad
  \text{\samp{LL}}\,\heq\,L_{\mathrm{L}}\,,\quad
  \text{\samp{eR}}\,\heq\,e_{\mathrm{R}}
\end{equation}
and their Hermitian conjugate. Their representations w.r.t.\ the Lorentz and
the \sm\ gauge group are defined in lines~24--53 of the model file.%
\footnote{ In the supplementary model files, the electromagnetic charge $Q$ is
defined by the relation $Q = I_3 + Y$ where $I_3$ and $Y$ are the \nth{3}
component of weak-isospin and the $U(1)$-hypercharge, respectively.  }

In principle, the second and third generation of fermions could be implemented
as separate copies of \cref{eq:sm:fermions}. More conveniently though, one may
add the entry \samp{generations:~3} to every fermion declaration, see
\cref{lst:sm}.  By using this option, \autoeft\ will associate a generation
index with these fields, which leads to a much more compact form of the
output, of course. Note that, even though the sum over generation indices is
not carried out explicitly in this case, the output does depend on the actual
number of generations. This is because the external and internal symmetries
may induce redundancies which depend on this number (see \citere{Fonseca:2019yya} for
details).

To complete the \sm, the complex Higgs doublet $H\heq$~\samp{H} must be included as well.
This is simply done by defining it as an $SU(2)$ doublet and assigning an appropriate hypercharge, see lines~54--57 of \cref{lst:sm}.

The entire model file for the \sm\ with three generations of fermions is then
given by \cref{lst:sm}.\footnote{As supplementary material, we supply the model file
\file{sm.yml}. Besides the information displayed in \cref{lst:sm}, this file
contains additional keywords which, however, only affect the \LaTeX\ markup of
the operators.  For consistency with other literature and
\citere{Harlander:2023psl}, we also supply the model file
\file{all-left\_sm.yml} that defines the fields in the all-left chirality
convention (cf.~\cref{sssec:qed}).  }
\begin{lstlisting}[caption={\sm\ model
    file.},label={lst:sm},style=yamlModel,float]
# AutoEFT model file
name: SMEFT
description: Standard Model Effective Field Theory

symmetries:
  sun_groups:
    SU3: {N: 3}
    SU2: {N: 2}
  u1_groups:
    U1: {}

fields:
  GL:
    representations:
      Lorentz: -1
      SU3: [2,1]
  WL:
    representations:
      Lorentz: -1
      SU2: [2]
  BL:
    representations:
      Lorentz: -1
  QL:
    representations:
      Lorentz: -1/2
      SU3: [1]
      SU2: [1]
      U1: 1/6
    generations: 3
  uR:
    representations:
      Lorentz: 1/2
      SU3: [1]
      U1: 2/3
    generations: 3
  dR:
    representations:
      Lorentz: 1/2
      SU3: [1]
      U1: -1/3
    generations: 3
  LL:
    representations:
      Lorentz: -1/2
      SU2: [1]
      U1: -1/2
    generations: 3
  eR:
    representations:
      Lorentz: 1/2
      U1: -1
    generations: 3
  H:
    representations:
      SU2: [1]
      U1: 1/2
\end{lstlisting}

%- }}}
%- }}}
%- {{{ subsection{Extended Models}

\subsection{Extended Models}\label{ssec:custom-models}
After reading \cref{ssec:description,ssec:examples}, and optionally consulting
\labelcref{anchor:Model}, the user should be able to assemble custom model
files from scratch.  However, \autoeft\ offers an alternative approach of
creating model files using the \code{sample-model} command.  Running this
command will print the content of a predefined \sm\ model file to the standard
output (e.g., the terminal).  Therefore, a custom model can also be obtained by
running the command
\begin{Texinfoindented}
\begin{Texinfopreformatted}%
\ttfamily autoeft sample-model > custom.yml
\end{Texinfopreformatted}
\end{Texinfoindented}
and subsequently modifying the newly created file \file{custom.yml} as
desired. Alternatively, the user may base the custom model on one of the
sample model files supplied with this paper. In the following, we consider
specific examples for extending the \sm\ as the low-energy theory.

%- {{{ subsubsection{Additional Particles}

\subsubsection{Additional Particles}\label{sssec:new_particles}

\citere{Banerjee:2020jun} defines a list of possible extensions of the \sm\ by
adding new particles. In order to illustrate the simplicity of preparing a
specific model file for \autoeft, we explicitly describe the necessary
modifications of the \sm\ model file for all examples provided in this paper.
Each model file can also be found in the supplementary material of this
paper, or in the \autoeft\ repository, see \cref{foot:repo}. It allows one to reconstruct the operator bases provided in
\citere{Banerjee:2020jun}, and to extend them to higher mass dimension.

Quite in general, new particles can be included in the \eft\ construction by
adding new entries under the keyword \code{fields} and assigning them
appropriate representations of the existing symmetry groups.  In the following
examples, we only show the lines that need to be added to the very end of the
default model file produced by the \code{sample-model} command. Following
\citere{Banerjee:2020jun} and adopting their notation, let us first consider
the addition of uncolored particles.

A scalar $\delta^+\heq\,$\samp{del} which only carries one unit of the hyper charge and
otherwise transforms as a singlet can be implemented as:
\begin{lstlisting}[linerange={81-83},firstnumber=79,style=yamlModel]
  del:
    representations:
      U1: 1
\end{lstlisting}  %chktex 8
for example. Of course, other (rational) values of the hyper charge can be
incorporated in an analogous way. For example, the doubly charged scalar named
$\rho^{++}\heq\,$\samp{rho} in \citere{Banerjee:2020jun} is obtained from:
\begin{lstlisting}[linerange={81-83},firstnumber=79,style=yamlModel]
  rho:
    representations:
      U1: 2
\end{lstlisting}  %chktex 8
Similarly, the complex scalar $SU(2)$-triplet $\Delta\heq\,$\samp{Del}
can be added as:
\begin{lstlisting}[linerange={81-84},firstnumber=79,style=yamlModel]
  Del:
    representations:
      SU2: [2]
      U1: 1
\end{lstlisting}  %chktex 8
and the left-handed fermion triplet $\Sigma\heq\,$\samp{Sig} is defined as:
\begin{lstlisting}[linerange={81-84},firstnumber=79,style=yamlModel]
  Sig:
    representations:
      Lorentz: -1/2
      SU2: [2]
\end{lstlisting}  %chktex 8
For vector-like leptons of various charges
($V_{\mathrm{L,R}},E_{\mathrm{L,R}},N_{\mathrm{L,R}}$)$\heq\,$(\samp{VL},\samp{VR},
\ldots), one also needs to
define the right-handed components:
\begin{lstlisting}[linerange={81-86,88-93,95-99,101-105,107-110,112-115},  %chktex 8
  firstnumber=79,style=yamlModel]
  VL:
    representations:
      Lorentz: -1/2
      SU2: [1]
      U1: -1/2
    generations: 3
  VR:
    representations:
      Lorentz: 1/2
      SU2: [1]
      U1: -1/2
    generations: 3
  EL:
    representations:
      Lorentz: -1/2
      U1: -1
    generations: 3
  ER:
    representations:
      Lorentz: 1/2
      U1: -1
    generations: 3
  NL:
    representations:
      Lorentz: -1/2
    generations: 3
  NR:
    representations:
      Lorentz: 1/2
    generations: 3
\end{lstlisting} Finally, also higher
representations of the gauge group can be accounted for. For example, the
scalar $SU(2)$-quadruplet $\Theta\heq\,$\samp{The} is given by:
\begin{lstlisting}[linerange={81-84},firstnumber=79,style=yamlModel]
  The:
    representations:
      SU2: [3]
      U1: 3/2
\end{lstlisting}  %chktex 8

New \emph{colored} particles can be included in exactly the same way by
assigning appropriate $SU(3)$ representations.  Again, we only show the lines
that need to be added to the very end of the default model file. In
particular, the various versions of lepto-quarks defined in
\citere{Banerjee:2020jun} can be implemented as:
\begin{description}
\item{Lepto-Quark ($\chi_1\heq\,$\samp{chi1})}
\begin{lstlisting}[linerange={81-85},firstnumber=79,style=yamlModel]
  chi1:
    representations:
      SU3: [1]
      SU2: [1]
      U1: 1/6
\end{lstlisting}  %chktex 8
\item{Lepto-Quark ($\varphi_1\heq\,$\samp{phi1})}
\begin{lstlisting}[linerange={81-84},firstnumber=79,style=yamlModel]
  phi1:
    representations:
      SU3: [1]
      U1: 2/3
\end{lstlisting}  %chktex 8
\item{Lepto-Quark ($\chi_2\heq\,$\samp{chi2})}
\begin{lstlisting}[linerange={81-85},firstnumber=79,style=yamlModel]
  chi2:
    representations:
      SU3: [1]
      SU2: [1]
      U1: 7/6
\end{lstlisting}  %chktex 8
\item{Lepto-Quark ($\varphi_2\heq\,$\samp{phi2})}
\begin{lstlisting}[linerange={81-84},firstnumber=79,style=yamlModel]
  phi2:
    representations:
      SU3: [1]
      U1: -1/3
\end{lstlisting}  %chktex 8
\end{description}

%- }}}
%- {{{ subsubsection{Additional \texorpdfstring{$U(1)$}

\subsubsection{Additional Gauge Symmetries}%
\label{sssec:extension-by-u-1-gauge-symmetries}%chktex 36
Additional gauge groups can be added by simply including their definition
under the keyword \code{symmetries}.  In the following example, there are two
new abelian gauge groups ${U(1)}^\prime$ and ${U(1)}^{\prime\prime}$,
extending the \sm\ gauge group.  Their respective gauge bosons are denoted by
$X$ and $Y$ (corresponding to \samp{XL}, \samp{YL} in the model file, plus the
automatically included conjugate fields).  In addition, global
symmetries---like baryon- and lepton-number conservation---can be added in
exactly the same way, with the only difference that there are no associated
gauge bosons.  In this example, each fermion gets assigned a specific baryon
and lepton number and the resulting operators must conserve the total numbers
exactly. Using the optional keys \code{violation} and \code{residual}, it
would also be possible to allow for a certain degree of violation of the
global $U(1)$ symmetries, see \labelcref{anchor:Model}.

The entire model file is displayed in \cref{lst:u1u1sm}, including the \code{tex}, \code{tex\_hc}, and \code{indices} keys that tell \autoeft\ how to
represent the symmetries, fields, and indices in \LaTeX\ format; see \labelcref{anchor:Model}.
New non-abelian gauge groups can be added in close analogy to the procedure
described above.
\begin{lstlisting}[caption={${U(1)}^\prime\times{U(1)}^{\prime\prime}$ extended
    model file.},label=lst:u1u1sm,style=yamlModel]
# AutoEFT model file
name: U(1)-U(1)-SMEFT
description: U(1)' x U(1)'' extended Standard Model Effective Field Theory

symmetries:
  lorentz_group:
    tex: SO^+(1,3)
  sun_groups:
    SU3:
      N: 3
      tex: SU(3)
      indices: [a,b,c,d,e,f,g,h]
    SU2:
      N: 2
      tex: SU(2)
      indices: [i,j,k,l,m,n,p,q]
  u1_groups:
    U1:
      tex: U(1)
    U1p:
      tex: U(1)^\prime
    U1pp:
      tex: U(1)^{\prime\prime}
    Bno: {}
    Lno: {}

fields:
  GL:
    representations:
      Lorentz: -1
      SU3: [2,1]
    tex: G_{\mathrm{L}}
    tex_hc: G_{\mathrm{R}}
  WL:
    representations:
      Lorentz: -1
      SU2: [2]
    tex: W_{\mathrm{L}}
    tex_hc: W_{\mathrm{R}}
  BL:
    representations:
      Lorentz: -1
    tex: B_{\mathrm{L}}
    tex_hc: B_{\mathrm{R}}
  XL:
    representations:
      Lorentz: -1
    tex: X_{\mathrm{L}}
    tex_hc: X_{\mathrm{R}}
  YL:
    representations:
      Lorentz: -1
    tex: Y_{\mathrm{L}}
    tex_hc: Y_{\mathrm{R}}
  QL:
    representations:
      Lorentz: -1/2
      SU3: [1]
      SU2: [1]
      U1: 1/6
      Bno: 1/3
    generations: 3
    tex: Q_{\mathrm{L}}
  uR:
    representations:
      Lorentz: 1/2
      SU3: [1]
      U1: 2/3
      Bno: 1/3
    generations: 3
    tex: u_{\mathrm{R}}
  dR:
    representations:
      Lorentz: 1/2
      SU3: [1]
      U1: -1/3
      Bno: 1/3
    generations: 3
    tex: d_{\mathrm{R}}
  LL:
    representations:
      Lorentz: -1/2
      SU2: [1]
      U1: -1/2
      Lno: -1
    generations: 3
    tex: L_{\mathrm{L}}
  eR:
    representations:
      Lorentz: 1/2
      U1: -1
      Lno: -1
    generations: 3
    tex: e_{\mathrm{R}}
  H:
    representations:
      SU2: [1]
      U1: 1/2
    tex: H
\end{lstlisting}

%- }}}
%- }}}
%- {{{ subsection{MFV Model}

\subsection{MFV Model}\label{ssec:mfv-model}
Instead of considering the three generations of fermions as independent
entities, one can also introduce so-called flavor symmetries.  In these
models, the approximate flavor symmetry of the \sm---which is only broken by
the Yukawa sector---is also imposed on the \eft.  A prominent example is
\mfv~\cite{Chivukula:1987py,DAmbrosio:2002vsn}, which introduces a global
${U(3)}^5 \sim {U(3)}_{Q} \times{U(3)}_{u} \times{U(3)}_{d} \times{U(3)}_{L}
\times{U(3)}_{e}$ flavor symmetry.  Although \autoeft\ does not support $U(N)$
symmetries directly, there is a Lie algebra isomorphism to $SU(N) \times U(1)$.
Hence, \mfv\ is realized by assigning an
${SU(3)}_f\heq\,$\samp{SU3<f>} fundamental representation and a
${U(1)}_f\heq\,$\samp{U1<f>}
(\code{<f>}$\in\{$\code{q},\code{u},\code{d},\code{l},\code{e}$\}$) charge of
unity to every fermion:
\begin{equation}
\begin{aligned}
	Q \sim \ydiagram{1}_{{SU(3)}_Q} \otimes 1_{{U(1)}_Q}
	\,,\quad
	u \sim \ydiagram{1}_{{SU(3)}_u} &\otimes 1_{{U(1)}_u}
	\,,\quad
	d \sim \ydiagram{1}_{{SU(3)}_d} \otimes 1_{{U(1)}_d}
	\,,\\[1em]
	L \sim \ydiagram{1}_{{SU(3)}_L} \otimes 1_{{U(1)}_L}
	\,,&\quad
	e \sim \ydiagram{1}_{{SU(3)}_e} \otimes 1_{{U(1)}_e}
	\,.
\end{aligned}
\end{equation}
\begin{lstlisting}[caption={\mfv\ model
    file.},label=lst:mfv,style=yamlModel]
# AutoEFT model file
name: MFV-SMEFT
description: Minimal Flavor Violation Standard Model Effective Field Theory

symmetries:
  sun_groups:
    SU3: {N: 3}
    SU2: {N: 2}
    SU3q: {N: 3}
    SU3u: {N: 3}
    SU3d: {N: 3}
    SU3l: {N: 3}
    SU3e: {N: 3}
  u1_groups: {U1: {}, U1q: {}, U1u: {}, U1d: {}, U1l: {}, U1e: {}}

fields:
  GL:
    representations: {Lorentz: -1, SU3: [2,1]}
  WL:
    representations: {Lorentz: -1, SU2: [2]}
  BL:
    representations: {Lorentz: -1}
  QL:
    representations: {Lorentz: -1/2, SU3: [1], SU2: [1], U1: 1/6, SU3q: [1], U1q: 1}
  uR:
    representations: {Lorentz: 1/2, SU3: [1], U1: 2/3, SU3u: [1], U1u: 1}
  dR:
    representations: {Lorentz: 1/2, SU3: [1], U1: -1/3, SU3d: [1], U1d: 1}
  LL:
    representations: {Lorentz: -1/2, SU2: [1], U1: -1/2, SU3l: [1], U1l: 1}
  eR:
    representations: {Lorentz: 1/2, U1: -1, SU3e: [1], U1e: 1}
  H:
    representations: {SU2: [1], U1: 1/2}
\end{lstlisting}
Since now every fermion carries a fundamental ${SU(3)}_f$ index, one must remove the entry
\samp{generations:~3} of \cref{lst:sm} from all fermion
declarations.
The entire model file encoding \mfv\ is shown in \cref{lst:mfv}.
It can be used to
construct the leading (i.e.,~flavor symmetric) terms in the \mfv\ \eft\ basis.%
\footnote{
In principle, it would be possible to include the Yukawa couplings as spurion fields---also transforming under the flavor symmetry.
This would allow to construct the \mfv\ \eft\ basis beyond the leading terms.
However, the Yukawa couplings are dimensionless and should instead be expanded by some other small quantity.
Such a declaration is not included in the model file specifications yet, but we intend to implement this feature in the next release of \autoeft.
}
Of course, other realizations of flavor symmetry can be implemented in a similar fashion.
For example, \citeres{Faroughy:2020ina,Greljo:2022cah,Greljo:2023adz} examine
various flavor symmetries in an \eft\ context.

%- }}}
%- }}}
%- {{{ section{Constructing Operators}

\section{Constructing Operators}\label{sec:constructing-operators}

%- {{{ subsection{Running \autoeft}

\subsection{Running \autoeft}\label{sec:running}

Given a valid model file, \autoeft\ can be used to construct an \eft\ basis
for a certain mass dimension.  For example, to construct the \smeft\ dimension-six
operators, run the command:%
\begin{Texinfoindented}
\begin{Texinfopreformatted}%
\ttfamily autoeft construct sm.yml 6
\end{Texinfopreformatted}
\end{Texinfoindented}
where \file{sm.yml} denotes the model file of \cref{lst:sm}\@.  \autoeft\ will
first display a disclaimer followed by a summary of the loaded model.  The
summary includes the name and description of the model as well as a table
containing all fields of the model, including the automatically generated
conjugate fields.  The table can be used to verify that the model file has
been loaded correctly and the field representations are set up as desired.
Afterwards, the operator construction starts and the number of families,
types, terms, and operators is displayed in a live preview (see
\labelcref{anchor:Vocabulary} and \citere{Harlander:2023psl} for the meaning
of these expressions).  After the operator construction is finished,
\autoeft\ terminates and returns to the shell prompt.
During each run, \autoeft\ writes a \emph{log file} called \file{autoeft.log} to the current working directory, capturing the console output.

During the construction, \autoeft\ creates the output directory
\file{efts/sm-eft/6/} in the current working directory. The substring
\samp{sm} is derived from the name of the model file \file{sm.yml}, and
\samp{6} is the requested mass dimension. All output files of \autoeft\ will
be written into this directory or its subdirectories.
If during the construction an operator type which is already present in the
output directory is encountered, \autoeft\ will skip the construction of this particular
type.\footnote{Unless the \code{{-}{-}overwrite} flag is set; see \labelcref{anchor:Construct}.}

The operator basis itself is written into the subdirectory \file{basis/}.
This directory always contains the file \file{model.json}, serving as a
reference to the model used during the construction, and the hidden file \file{.autoeft} containing metadata of the generation.  All constructed operator
files of a given \code{family} and \code{type}
(cf.~\labelcref{anchor:Vocabulary}) are included in further subdirectories of
the form \file{<N>/<family>/<type>.yml}, where \code{N} denotes the total
number of fields in the operator.  The format of the operator files is
explained in the next section.

A detailed description of all command-line options of the
\code{construct}~(short:~\code{c}) command can be found in
\labelcref{anchor:Construct}.  Here, we only mention the optional
\code{{-}{-}select}~(short:~\code{-s}) and
\code{{-}{-}ignore}~(short:~\code{-i}) options, which are particularly useful
if only a specific subset of operators should be constructed.  For example, to
only construct dimension-six operators containing exactly two Higgs doublets,
run the command:
\begin{Texinfoindented}
\begin{Texinfopreformatted}%
  \ttfamily autoeft c sm.yml 6 -s "\{H: 2, H+: 0\}" -s "\{H: 0, H+: 2\}" \textbackslash
  \phantom{autoeft c sm.yml 6} -s "\{H: 1, H+: 1\}"  %chktex 18
\end{Texinfopreformatted}
\end{Texinfoindented}
On the other hand, the command
\begin{Texinfoindented}
\begin{Texinfopreformatted}%
\ttfamily autoeft c sm.yml 6 -i "\{GL: +\}" -i "\{GL+: +\}"  %chktex 18
\end{Texinfopreformatted}
\end{Texinfoindented}
will exclude all operators containing gluons.  The \code{-s} and \code{-i}
options can be combined, of course, whereupon the latter overrides the former
in case of conflicts.

After a successful run, \autoeft\ writes the file \file{stats.yml} to the output directory, containing the total number of families, types, terms, and operators in the basis.
These numbers can also be obtained using the \code{count} command; see \labelcref{anchor:Count}.

%- }}}
%- {{{ subsection{Output Format}

\subsection{Output Format}\label{ssec:operator-basis}

The \emph{operator files} contain all information needed to reconstruct the
\eft\ basis type-by-type.  Here, we demonstrate how their content can be
interpreted using the \smeft\ operator type $L^1_{\mathrm{L}}Q^3_{\mathrm{L}}$
as an example.  The entire operator file, named \code{1LL\_3QL.yml}, is
displayed in \cref{lst:1L3Q}.\footnote{In the supplemental material accompanying
\citere{Harlander:2023psl}, which adopts the all-left notation for the fields,
the corresponding file is named \code{1L\_3Q.yml}.}
\begin{lstlisting}[caption={$L^1_{\mathrm{L}}Q^3_{\mathrm{L}}$ operator
    file.},label={lst:1L3Q},style=yamlOperator]
# '1LL_3QL.yml' generated by AutoEFT 1.0.0
version: 1.0.0
type:
- {LL: 1, QL: 3}
- complex
generations: {LL: 3, QL: 3}
n_terms: 3
n_operators: 57
invariants:
  Lorentz:
    O(Lorentz,1): +eps(1_1,3_1)*eps(2_1,4_1) * LL(1_1)*QL(2_1)*QL(3_1)*QL(4_1)
    O(Lorentz,2): +eps(1_1,2_1)*eps(3_1,4_1) * LL(1_1)*QL(2_1)*QL(3_1)*QL(4_1)
  SU3:
    O(SU3,1): +eps(2_1,3_1,4_1) * LL*QL(2_1)*QL(3_1)*QL(4_1)
  SU2:
    O(SU2,1): +eps(1_1,3_1)*eps(2_1,4_1) * LL(1_1)*QL(2_1)*QL(3_1)*QL(4_1)
    O(SU2,2): +eps(1_1,2_1)*eps(3_1,4_1) * LL(1_1)*QL(2_1)*QL(3_1)*QL(4_1)
permutation_symmetries:
- vector: Lorentz * SU3 * SU2
- symmetry: {LL: [1], QL: [1, 1, 1]}
  n_terms: 1
  n_operators: 3
  matrix: |-
    [ 0 -1  1  0]
- symmetry: {LL: [1], QL: [2, 1]}
  n_terms: 1
  n_operators: 24
  matrix: |-
    [-1  2  2 -1]
- symmetry: {LL: [1], QL: [3]}
  n_terms: 1
  n_operators: 30
  matrix: |-
    [ 2 -1 -1  2]
\end{lstlisting}

A summary of all keywords appearing in the operator files is included in \labelcref{anchor:Output}.
For this particular example, they can be interpreted in the following way:
\begin{description}[style=nextline]
\item[\code{version}:]
The version of \autoeft\ that was used to produce the output file.

\item[\code{type}:] The first entry denotes the operator type, in this example
  $L^1_{\mathrm{L}}Q^3_{\mathrm{L}}$.  The second entry states that this type
  is \samp{complex}, meaning there is a distinct Hermitian conjugate type
  (which is contained in \code{1LL+\_3QL+.yml}).

\item[\code{generations}:]
For reference, the number of generations for each field is also displayed in the operator files.
In the present case, the file was generated for three generations of leptons and quarks.

\item[\code{n\_terms}:] The total number of operators with independent Lorentz
  and internal index contractions and definite permutation symmetry of the
  repeated fields (i.e.\ fields which differ at most in their generation
  index).  It does not take into account the different generations though.  In
  this example, the generation indices of the quarks can be decomposed into
  totally anti-symmetric $[1,1,1]$, mixed symmetric $[2,1]$, and totally symmetric $[3]$ tensors.

\item[\code{n\_operators}:]
The total number of independent operators, taking into account the independent values the generation indices can assume.
Here, there are $3 \cdot (1 + 8 + 10) = 57$ independent combinations of the $L_{\mathrm{L}}$ and $Q_{\mathrm{L}}$ generations.

\item[\code{invariants}:]
The invariant contractions are given by:
\begin{equation}\label{eq:contractions}
  \begin{aligned}
    \mathcal{O}^{\text{Lorentz}}_{1} &=
	\epsilon^{\alpha\gamma}\,\epsilon^{\beta\delta}\,
	{L_\mathrm{L}}_{\alpha}\,{Q_\mathrm{L}}_{\beta}\,{Q_\mathrm{L}}_{\gamma}\,{Q_\mathrm{L}}_{\delta}
	\,,\\[.5em]
    \mathcal{O}^{\text{Lorentz}}_{2} &=
	\epsilon^{\alpha\beta}\,\epsilon^{\gamma\delta}\,
	{L_\mathrm{L}}_{\alpha}\,{Q_\mathrm{L}}_{\beta}\,{Q_\mathrm{L}}_{\gamma}\,{Q_\mathrm{L}}_{\delta}
	\,,\\[1.5em]
    \mathcal{O}^{SU(3)}_{1} &=
	\epsilon^{bcd}\,
	{L_\mathrm{L}}\,{Q_\mathrm{L}}_{b}\,{Q_\mathrm{L}}_{c}\,{Q_\mathrm{L}}_{d}
	\,,\\[1.5em]
	\mathcal{O}^{SU(2)}_{1} &=
	\epsilon^{ik}\,\epsilon^{jl}\,
	{L_\mathrm{L}}_{i}\,{Q_\mathrm{L}}_{j}\,{Q_\mathrm{L}}_{k}\,{Q_\mathrm{L}}_{l}
	\,,\\[.5em]
	\mathcal{O}^{SU(2)}_{2} &=
	\epsilon^{ij}\,\epsilon^{kl}\,
	{L_\mathrm{L}}_{i}\,{Q_\mathrm{L}}_{j}\,{Q_\mathrm{L}}_{k}\,{Q_\mathrm{L}}_{l}
	\,,
  \end{aligned}
\end{equation}
  where only the relevant set of indices is displayed in each case.
\item[\code{permutation\_symmetries}:]
The first entry always denotes the order of the tensor product.
In this case, the combination is given by
\begin{equation}
	\mathtt{vector:}\quad
	\vec{\mathcal{O}} \equiv \mathcal{O}^{\text{Lorentz}} \otimes \mathcal{O}^{SU(3)} \otimes \mathcal{O}^{SU(2)} =
	\begin{psmallmatrix}
	   \mathcal{O}^{\text{Lorentz}}_1\,\otimes\, \mathcal{O}^{SU(3)}_1\,\otimes\, \mathcal{O}^{SU(2)}_1
		\\[.3em]
	   \mathcal{O}^{\text{Lorentz}}_1\,\otimes\, \mathcal{O}^{SU(3)}_1\,\otimes\, \mathcal{O}^{SU(2)}_2
		\\[.3em]
	   \mathcal{O}^{\text{Lorentz}}_2\,\otimes\, \mathcal{O}^{SU(3)}_1\,\otimes\, \mathcal{O}^{SU(2)}_1
		\\[.3em]
	   \mathcal{O}^{\text{Lorentz}}_2\,\otimes\, \mathcal{O}^{SU(3)}_1\,\otimes\, \mathcal{O}^{SU(2)}_2
	\end{psmallmatrix}
	\,.
\end{equation}
The first element of this vector is to be read as
\begin{equation}\label{eq:operator:erst}
  \begin{aligned}
   & \mathcal{O}^{\text{Lorentz}}_1\,\otimes\, \mathcal{O}^{SU(3)}_1\,\otimes\,
    \mathcal{O}^{SU(2)}_1 =\\&\qquad=
    \epsilon^{\alpha\gamma}\,\epsilon^{\beta\delta}\,
	\epsilon^{ik}\,\epsilon^{jl}\,
	\epsilon^{bcd}\,
	        {L_\mathrm{L}^w}_{\alpha i}\,{Q^x_\mathrm{L}}_{\beta b j}
            \,{Q^y_\mathrm{L}}_{\gamma c k}\,{Q^z_\mathrm{L}}_{\delta d l}\,,
  \end{aligned}
\end{equation}
for example, with generation indices $w,x,y,z\in\{1,2,3\}$, while the
other indices are those of \cref{eq:contractions}.

The choice of generation indices in the repeated fields is not arbitrary
though. The remaining entries take this into account via the permutation
symmetries of the repeated fields and the associated linearly independent
combinations of the invariant contractions.  In this case, there are three
distinct permutation symmetries of the quark generation indices:
\begin{itemize}
\item $\lambda_{L_{\mathrm{L}}} \sim [1]\quad\&\quad\lambda_{Q_{\mathrm{L}}} \sim [1,1,1]$
\begin{equation}
    \mathtt{matrix:}\quad
    \mathcal{K}^{[1],[1,1,1]}\equiv
    \begin{pmatrix}
        0 & -1 & 1 & 0
    \end{pmatrix}\,.
\end{equation}

\item $\lambda_{L_{\mathrm{L}}} \sim [1]\quad\&\quad\lambda_{Q_{\mathrm{L}}} \sim [2,1]$
\begin{equation}
    \mathtt{matrix:}\quad
	\mathcal{K}^{[1],[2,1]}\equiv
    \begin{pmatrix}
        -1 & 2 & 2 & -1
    \end{pmatrix}\,.
\end{equation}

\item $\lambda_{L_{\mathrm{L}}} \sim [1]\quad\&\quad\lambda_{Q_{\mathrm{L}}} \sim [3]$
\begin{equation}
    \mathtt{matrix:}\quad
	\mathcal{K}^{[1],[3]}\equiv
    \begin{pmatrix}
        2 & -1 & -1 & 2
    \end{pmatrix}\,.
\end{equation}
\end{itemize}
\end{description}

Combining the information of the model file, one arrives at three independent
terms, each with definite permutation symmetry. The first one is
\begin{equation}\label{eq:expanded}
  \begin{aligned}
	\mathcal{O}^{[1],[1,1,1]}&\equiv\mathcal{K}^{[1],[1,1,1]}\cdot\vec{\mathcal{O}}\\
	&=
	-\,\mathcal{O}^{\text{Lorentz}}_1\,\otimes\, \mathcal{O}^{SU(3)}_1\,\otimes\, \mathcal{O}^{SU(2)}_2
	+\,\mathcal{O}^{\text{Lorentz}}_2\,\otimes\, \mathcal{O}^{SU(3)}_1\,\otimes\, \mathcal{O}^{SU(2)}_1
	\,.
  \end{aligned}
\end{equation}
While the generation index $w\in\{1,2,3\}$ for the lepton can be any of these
three values, the quark generation indices can only assume a single
combination of values in this case; one choice is $(x,y,z)=(1,2,3)$, for
example. Thus, $\mathcal{O}^{[1],[1,1,1]}$ represents $3\cdot 1=3$ different
operators (see line~22 of \cref{lst:1L3Q}), if the generation of the fields is
taken into account.

The second term is
\begin{equation}
  \begin{aligned}
\mathcal{O}^{[1],[2,1]}&\equiv\mathcal{K}^{[1],[2,1]}\cdot\vec{\mathcal{O}}\\
	&=
    -\,\mathcal{O}^{\text{Lorentz}}_1\,\otimes\, \mathcal{O}^{SU(3)}_1\,\otimes\, \mathcal{O}^{SU(2)}_1
    +2\,\mathcal{O}^{\text{Lorentz}}_1\,\otimes\, \mathcal{O}^{SU(3)}_1\,\otimes\, \mathcal{O}^{SU(2)}_2
	\\&\quad
	+2\,\mathcal{O}^{\text{Lorentz}}_2\,\otimes\, \mathcal{O}^{SU(3)}_1\,\otimes\, \mathcal{O}^{SU(2)}_1
    -\,\mathcal{O}^{\text{Lorentz}}_2\,\otimes\, \mathcal{O}^{SU(3)}_1\,\otimes\, \mathcal{O}^{SU(2)}_2
    \,.
  \end{aligned}
\end{equation}
For this permutation symmetry, there are eight independent combinations of the
quark generation indices; one may choose them to be\footnote{They can be
determined from the associated semi-standard Young tableaux
$\ytableaushort{xy,z}$; see \citere{Harlander:2023psl} for details.}
\begin{equation}\label{eq:operator:dike}
  \begin{aligned}
    (x,y,z) \in \{& (1,1,2),(1,1,3),(1,2,2),(1,2,3),(1,3,2),\\&(1,3,3),
    (2,2,3),(2,3,3) \}\,.
  \end{aligned}
\end{equation}
Again taking into account the multiplicity of the lepton generations, this
term represents $3\cdot 8=24$ operators (see line~27).

The third term is
\begin{equation}
  \begin{aligned}
	\mathcal{O}^{[1],[3]}&\equiv\mathcal{K}^{[1],[3]}\cdot\vec{\mathcal{O}}\\
	&=
    2\,\mathcal{O}^{\text{Lorentz}}_1\,\otimes\, \mathcal{O}^{SU(3)}_1\,\otimes\, \mathcal{O}^{SU(2)}_1
    -\,\mathcal{O}^{\text{Lorentz}}_1\,\otimes\, \mathcal{O}^{SU(3)}_1\,\otimes\, \mathcal{O}^{SU(2)}_2
	\\&\quad
    -\,\mathcal{O}^{\text{Lorentz}}_2\,\otimes\, \mathcal{O}^{SU(3)}_1\,\otimes\, \mathcal{O}^{SU(2)}_1
    +2\,\mathcal{O}^{\text{Lorentz}}_2\,\otimes\, \mathcal{O}^{SU(3)}_1\,\otimes\, \mathcal{O}^{SU(2)}_2
	\,,
  \end{aligned}
\end{equation}
where one can choose the following ten combinations of the quark generation
indices:
\begin{equation}\label{eq:operator:iota}
  \begin{aligned}
    (x,y,z) \in \{& (1,1,1),(1,1,2),(1,1,3),(1,2,2),(1,2,3),\\ &(1,3,3),
    (2,2,2),(2,2,3),(2,3,3),(3,3,3) \}\,,
  \end{aligned}
\end{equation}
and therefore this term represents $3\cdot 10=30$ operators (see line~32).

More examples and detailed descriptions of the output format can be found in
\citere{Harlander:2023psl}.
See also \cref{ssec:loading-operator-files} for an example on how \autoeft\ can be used to perform this expansion automatically.

%- }}}
%- {{{ subsection{Counting SMEFT Operators}

\subsection{Counting SMEFT Operators}\label{ssec:counting-smeft-operators}

The number of families, types, terms, and operators
(cf.~\labelcref{anchor:Vocabulary}) can be obtained from an existing
basis using the \code{count}
command;\footnote{ See \labelcref{anchor:basis} for the definition of a valid
operator basis that can be processed by \autoeft.  } see \labelcref{anchor:Count}.
\autoeft\ also writes these numbers to
the file \code{stats.yml} after each basis construction, see
\cref{sec:running}.  In order to arrive at a well-defined number, operators
(families, types, terms) and their distinct conjugate version are counted
separately. This means that, for \smeft\ at mass dimension five,
\autoeft\ counts two operators (equaling the number of families, terms, and
types) if only one generation of leptons is taken into account: the Weinberg
operator $\sim L_\mathrm{L} L_\mathrm{L} HH$ and its conjugate $\sim H^\dagger
H^\dagger L_\mathrm{L}^\dagger L_\mathrm{L}^\dagger$. For three generations,
each of the two terms leads to six operators (permutation symmetry eliminates
three out of the nine possible combinations, cf.~\cref{ssec:operator-basis}).

For \smeft\ at mass dimension six, various different numbers can be found in
the literature, depending on the way of counting, or possible additional
symmetries that have been imposed. For example, assuming baryon-number
conservation, \autoeft\ generates 76 operators for a single generation of
fermions. This is in line with the 59 operators reported in
\citere{Grzadkowski:2010es} if one counts the 17 Hermitian conjugate operators
of $Q_{uG}$, $Q_{dG}$, $Q_{\varphi ud}$, $Q_{ledq}$, ${Q_{quqd}^{(1)}}$,
${Q_{quqd}^{(8)}}$, ${Q_{lequ}^{(1)}}$, ${Q_{lequ}^{(3)}}$, and $Q_{ij}$ with
$i\in\{e,u,d\}$, $j\in\{\varphi,W,B\}$, as independent degrees of freedom.
Relaxing the assumption of baryon-number conservation, \autoeft\ reports a
total of 84 operators, corresponding to the addition of the four baryon-number
violating operators $Q_{duq}$, $Q_{qqu}$, $Q_{qqq}$, and $Q_{duu}$, quoted in
\citere{Grzadkowski:2010es}, and their Hermitian conjugate versions.  In the
case of three fermion generations, \autoeft\ generates 2499 operators if
baryon-number conservation is imposed; otherwise the number increases to 3045.
Enforcing flavor conservation as described in \cref{ssec:mfv-model},
\autoeft\ only generates 47 operators at mass dimension six
(cf.~\citere{Greljo:2022cah}).

The number of operators calculated with the Hilbert
series~\cite{Henning:2015alf,Marinissen:2020jmb} matches the counting of
\autoeft\ for any mass dimension exactly, as we have verified for \smeft\ up
to mass dimension~12~\cite{Harlander:2023psl}.

%- }}}
%- {{{ subsection{Hermitian Conjugate Operators}

\subsection{Hermitian Conjugate Operators}\label{ssec:hermitian-conjugate-operators}
To write down a
\emph{real} \eft~Lagrangian, it is required to include the Hermitian conjugate
version of each term. Given a complex Wilson coefficient $\mathcal{C}$ and an operator
$\mathcal{O}$, this means that the Lagrangian must contain the sum
\begin{equation}\label{eq:real}
\mathcal{C}\mathcal{O} + \mathcal{C}^\ast\mathcal{O}^\dagger \in \mathcal{L} \,.
\end{equation}
If the operator $\mathcal{O}$ is (anti-)Hermitian, \cref{eq:real} simplifies
to $(\mathcal{C} \pm \mathcal{C}^\ast)\mathcal{O}$ so that only the real (imaginary) part of the
Wilson coefficient contributes to the Lagrangian, and the operator represents
a single degree of freedom.  If the operator is not (anti-)Hermitian, both
real and imaginary parts of the Wilson coefficient appear in the Lagrangian,
and the operator represents two degrees of freedom.  Of course, one could
choose a basis of only Hermitian operators (implying only real Wilson
coefficients) which is, however, not very practical for most applications.

\autoeft\ treats operators that are obtained by Hermitian conjugation as
independent degrees of freedom---except for (anti-)Hermitian operators. By
default, it thus includes the Hermitian conjugate operator types in the basis.
Note that the Hermitian conjugate operators of a given type are not
necessarily the same as the operators of the conjugate type, but rather linear
combinations thereof.  Future versions of \autoeft\ will provide the
functionality to obtain the Hermitian conjugate of individual operators,
rather than operator types.

\autoeft\ can be prevented from constructing the Hermitian conjugate types by
the command-line option \code{{-}{-}no\_hc}.  The user then needs to conjugate
each operator that is constructed by \autoeft\ and add it to the Lagrangian
with the complex conjugate Wilson coefficient, see \cref{eq:real}.

%- }}}
%- {{{ subsection{Limitations}

\subsection{Limitations of \autoeft}\label{ssec:limitations}

%- {{{ subsubsection{Computing Resources}

\subsubsection{Computing Resources}

The algorithm implemented in \autoeft\ works for an arbitrary mass dimension
$d$. However, the computational efforts increase exponentially with $d$. For
\smeft, for example, while the basis at mass dimension~10 could be constructed
within a few hours, dimension~12 took of the order of months. Currently, CPU
time is therefore probably the most severe limiting factor for going to even
higher mass dimension.

Other hardware limitations may arise from memory requirements, which mostly
come from algebraic operations when projecting the general Lorentz and $SU(N)$
tensors onto the tensor basis. Also the storage of the output files may exceed
the available hardware. In \citere{Harlander:2023psl}, we estimated the
required disk space for \smeft\ at mass dimension~26 to amount to about 1~PB.

%- }}}
%- {{{ subsubsection{Generators of the Symmetric Group}

\subsubsection{Generators of the Symmetric Group}\label{ssec:generators-of-the-symmetric-group}

The elimination of redundancies due to the occurrence of repeated fields in an
operator requires representation matrices of the symmetric
group $S_n$. Since these are independent of the specific \eft\ under
consideration, \autoeft\ comes with a hard-coded version of generator matrices
that are used to generate all required representation matrices up
to $n=9$, which is sufficient for problems with up to nine
repeated fields in an operator.\footnote{For \smeft, this would be sufficient
to generate the basis up to mass dimension~18.} If the construction of an
\eft\ requires a representation matrix which is not hard-coded, \autoeft\ will
terminate and request the missing matrices\@. \autoeft\ also provides a
functionality to (re-)compute these matrices, using the \code{generators}
command, see \labelcref{anchor:Commands}. Calculating them beyond $n=9$ is very
\abbrev{CPU} expensive though.

%- }}}

\subsubsection{Conceptual Limitations}

As indicated already in \cref{sec:preliminaries}, \autoeft\ is currently
limited to fields with spin 0, 1/2, 1, and 2, where the latter two must be
massless. Furthermore, internal symmetries must only be due to $U(1)$ or
$SU(N)$ groups.

The operator basis will be non-redundant on-shell, which means that operators
which are related by equations-of-motion and integration-by-parts identities
have been identified. For the purpose of renormalization, operators
proportional to equations-of-motion as well as gauge-variant operators are
required in general though, but currently these cannot be generated by
\autoeft.  In addition, \autoeft\ only constructs operators that mediate
proper interactions, meaning that any operator must be composed of at least
three fields.  Furthermore, \autoeft\ does not take into account evanescent
operators, as they may be required for calculations in dimensional
regularization (see, e.g., \citere{Herrlich:1994kh}).

%- }}}
%- }}}
%- {{{ section{Working With Operator Files}

\section{Working With Operator Files}\label{sec:operator-files}

%- {{{ intro:

Besides providing a command-line script for the operator basis construction,
\autoeft\ also serves as a \sage\ library to work with \eft\ operators.  In
this section, we illustrate some of the features available by importing
\autoeft\ into \sage, which will be extended further in future releases of
\autoeft.  We also introduce some additional command-line functionalities that
are not directly related to the operator construction but may be useful for
the user.

%- }}}
%- {{{ subsection{Loading Operator Files}

\subsection{Loading Operator Files}\label{ssec:loading-operator-files}

Once an operator basis is constructed using the \code{construct} command, it
is possible to load the operator files into \sage\ for further manipulation.
A \emph{valid} operator basis that can be processed by \autoeft\ is
represented by a directory containing the file \file{model.json} referencing
the model, a (hidden) file \file{.autoeft} containing metadata, and the
respective operator files in the format described in \cref{ssec:operator-basis}.%
\footnote{ Note that, while the files \file{model.json} and \file{.autoeft}
must be contained in the top-level directory of the basis, the operator files
can be structured in subdirectories.  The \code{get\_basis} function searches
all subdirectories for files with extension \samp{.yml} and loads them as
operator file.  } By default, the directories called \file{basis} created by
\autoeft\ are structured such that they can be directly loaded into \sage.
See also \labelcref{anchor:basis}.

Consider for example \cref{lst:load}, displaying how one can load the mass
dimension-six \smeft\ basis located at \file{efts/sm-eft/6/basis}
(cf.~line~5). The command lines in this listing can be entered into an
interactive \sage\ session, for example, or stored in a \code{<file>} first
and passed to sage via a shell command \samp{sage <file>}\@. \autoeft\ provides
the class \code{autoeft.io.basis.BasisFile} to interact with the basis stored
on the disk.  It must be initialized with the path pointing to the basis as
displayed in line~6.  Afterwards, the entire basis can be loaded into memory
with the \code{get\_basis()} function (cf.~line~7), returning a dictionary
that maps each operator type to the contents of the respective operator file.
For example, to access the operators of type
$L_{\mathrm{L}}^1Q_{\mathrm{L}}^3$ (see also \cref{ssec:operator-basis}), the
type can be identified by the mapping \samp{\{"LL":~1,~"QL":~3\}}, following
the same convention as the output files; see line~4 in \cref{lst:1L3Q}.  The
resulting object, assigned to the variable \samp{LQQQ} in line~9 of
\cref{lst:load}, is an instance of the
\code{autoeft.base.basis.OperatorInfoPermutation} class.  This class contains
all the information of the operator files as object properties, see for
example lines~10--14.  However, it also contains further properties and
functions.  For example, to obtain the actual terms of this type, the object
can be expanded using the \code{expanded()} function, returning a new object
that contains the terms as displayed in \cref{eq:expanded} (cf.~lines~16--22
of \cref{lst:load}).

\begin{lstlisting}[style=pythonSample,label=lst:load,caption={Example script loading an operator basis created by \autoeft. The operator type $L_{\mathrm{L}}^1Q_{\mathrm{L}}^3$ is selected and explicitly expanded into its terms as illustrated in \cref{ssec:operator-basis}.}]
from pathlib import Path

from autoeft.io.basis import BasisFile

basis_path = Path("efts/sm-eft/6/basis")
basis_file = BasisFile(basis_path)
basis = basis_file.get_basis()

LQQQ = basis[{"LL": 1, "QL": 3}]
print(LQQQ)
# LL(1) QL(3)

print(LQQQ.n_terms, LQQQ.n_operators, sep=" & ")
# 3 & 57

for term in LQQQ.expanded():
    print(term.symmetry)
    print(term)
    print(term.operators)
# {'LL': [1], 'QL': [1,1,1]}
# (-1) * eps(Lorentz_1_1,Lorentz_3_1)*eps(Lorentz_2_1,Lorentz_4_1)*eps(SU3_2_1,SU3_3_1,SU3_4_1)*eps(SU2_1_1,SU2_2_1)*eps(SU2_3_1,SU2_4_1) * LL(Lorentz_1_1;SU2_1_1)*QL(Lorentz_2_1;SU3_2_1;SU2_2_1)*QL(Lorentz_3_1;SU3_3_1;SU2_3_1)*QL(Lorentz_4_1;SU3_4_1;SU2_4_1) + (1) * eps(Lorentz_1_1,Lorentz_2_1)*eps(Lorentz_3_1,Lorentz_4_1)*eps(SU3_2_1,SU3_3_1,SU3_4_1)*eps(SU2_1_1,SU2_3_1)*eps(SU2_2_1,SU2_4_1) * LL(Lorentz_1_1;SU2_1_1)*QL(Lorentz_2_1;SU3_2_1;SU2_2_1)*QL(Lorentz_3_1;SU3_3_1;SU2_3_1)*QL(Lorentz_4_1;SU3_4_1;SU2_4_1)
# [(1, 1, 2, 3), (2, 1, 2, 3), (3, 1, 2, 3)]

# {'LL': [1], 'QL': [2,1]}
# (-1) * eps(Lorentz_1_1,Lorentz_3_1)*eps(Lorentz_2_1,Lorentz_4_1)*eps(SU3_2_1,SU3_3_1,SU3_4_1)*eps(SU2_1_1,SU2_3_1)*eps(SU2_2_1,SU2_4_1) * LL(Lorentz_1_1;SU2_1_1)*QL(Lorentz_2_1;SU3_2_1;SU2_2_1)*QL(Lorentz_3_1;SU3_3_1;SU2_3_1)*QL(Lorentz_4_1;SU3_4_1;SU2_4_1) + (2) * eps(Lorentz_1_1,Lorentz_3_1)*eps(Lorentz_2_1,Lorentz_4_1)*eps(SU3_2_1,SU3_3_1,SU3_4_1)*eps(SU2_1_1,SU2_2_1)*eps(SU2_3_1,SU2_4_1) * LL(Lorentz_1_1;SU2_1_1)*QL(Lorentz_2_1;SU3_2_1;SU2_2_1)*QL(Lorentz_3_1;SU3_3_1;SU2_3_1)*QL(Lorentz_4_1;SU3_4_1;SU2_4_1) + (2) * eps(Lorentz_1_1,Lorentz_2_1)*eps(Lorentz_3_1,Lorentz_4_1)*eps(SU3_2_1,SU3_3_1,SU3_4_1)*eps(SU2_1_1,SU2_3_1)*eps(SU2_2_1,SU2_4_1) * LL(Lorentz_1_1;SU2_1_1)*QL(Lorentz_2_1;SU3_2_1;SU2_2_1)*QL(Lorentz_3_1;SU3_3_1;SU2_3_1)*QL(Lorentz_4_1;SU3_4_1;SU2_4_1) + (-1) * eps(Lorentz_1_1,Lorentz_2_1)*eps(Lorentz_3_1,Lorentz_4_1)*eps(SU3_2_1,SU3_3_1,SU3_4_1)*eps(SU2_1_1,SU2_2_1)*eps(SU2_3_1,SU2_4_1) * LL(Lorentz_1_1;SU2_1_1)*QL(Lorentz_2_1;SU3_2_1;SU2_2_1)*QL(Lorentz_3_1;SU3_3_1;SU2_3_1)*QL(Lorentz_4_1;SU3_4_1;SU2_4_1)
# [(1, 1, 1, 2), (1, 1, 1, 3), (1, 1, 2, 2), (1, 1, 2, 3), (1, 1, 3, 2), (1, 1, 3, 3), (1, 2, 2, 3), (1, 2, 3, 3), (2, 1, 1, 2), (2, 1, 1, 3), (2, 1, 2, 2), (2, 1, 2, 3), (2, 1, 3, 2), (2, 1, 3, 3), (2, 2, 2, 3), (2, 2, 3, 3), (3, 1, 1, 2), (3, 1, 1, 3), (3, 1, 2, 2), (3, 1, 2, 3), (3, 1, 3, 2), (3, 1, 3, 3), (3, 2, 2, 3), (3, 2, 3, 3)]

# {'LL': [1], 'QL': [3]}
# (2) * eps(Lorentz_1_1,Lorentz_3_1)*eps(Lorentz_2_1,Lorentz_4_1)*eps(SU3_2_1,SU3_3_1,SU3_4_1)*eps(SU2_1_1,SU2_3_1)*eps(SU2_2_1,SU2_4_1) * LL(Lorentz_1_1;SU2_1_1)*QL(Lorentz_2_1;SU3_2_1;SU2_2_1)*QL(Lorentz_3_1;SU3_3_1;SU2_3_1)*QL(Lorentz_4_1;SU3_4_1;SU2_4_1) + (-1) * eps(Lorentz_1_1,Lorentz_3_1)*eps(Lorentz_2_1,Lorentz_4_1)*eps(SU3_2_1,SU3_3_1,SU3_4_1)*eps(SU2_1_1,SU2_2_1)*eps(SU2_3_1,SU2_4_1) * LL(Lorentz_1_1;SU2_1_1)*QL(Lorentz_2_1;SU3_2_1;SU2_2_1)*QL(Lorentz_3_1;SU3_3_1;SU2_3_1)*QL(Lorentz_4_1;SU3_4_1;SU2_4_1) + (-1) * eps(Lorentz_1_1,Lorentz_2_1)*eps(Lorentz_3_1,Lorentz_4_1)*eps(SU3_2_1,SU3_3_1,SU3_4_1)*eps(SU2_1_1,SU2_3_1)*eps(SU2_2_1,SU2_4_1) * LL(Lorentz_1_1;SU2_1_1)*QL(Lorentz_2_1;SU3_2_1;SU2_2_1)*QL(Lorentz_3_1;SU3_3_1;SU2_3_1)*QL(Lorentz_4_1;SU3_4_1;SU2_4_1) + (2) * eps(Lorentz_1_1,Lorentz_2_1)*eps(Lorentz_3_1,Lorentz_4_1)*eps(SU3_2_1,SU3_3_1,SU3_4_1)*eps(SU2_1_1,SU2_2_1)*eps(SU2_3_1,SU2_4_1) * LL(Lorentz_1_1;SU2_1_1)*QL(Lorentz_2_1;SU3_2_1;SU2_2_1)*QL(Lorentz_3_1;SU3_3_1;SU2_3_1)*QL(Lorentz_4_1;SU3_4_1;SU2_4_1)
# [(1, 1, 1, 1), (1, 1, 1, 2), (1, 1, 1, 3), (1, 1, 2, 2), (1, 1, 2, 3), (1, 1, 3, 3), (1, 2, 2, 2), (1, 2, 2, 3), (1, 2, 3, 3), (1, 3, 3, 3), (2, 1, 1, 1), (2, 1, 1, 2), (2, 1, 1, 3), (2, 1, 2, 2), (2, 1, 2, 3), (2, 1, 3, 3), (2, 2, 2, 2), (2, 2, 2, 3), (2, 2, 3, 3), (2, 3, 3, 3), (3, 1, 1, 1), (3, 1, 1, 2), (3, 1, 1, 3), (3, 1, 2, 2), (3, 1, 2, 3), (3, 1, 3, 3), (3, 2, 2, 2), (3, 2, 2, 3), (3, 2, 3, 3), (3, 3, 3, 3)]
\end{lstlisting}

%- }}}
%- {{{ subsection{LaTeX Output}

\subsection{\LaTeX\ Output}\label{ssec:latex-output}
\autoeft\ provides the \code{latex} (short:~\code{l}) command for the automatic \LaTeX\ markup of the constructed operators.
For example, to produce the \TeX\ files corresponding to the operator basis
located at \code{efts/sm-eft/6/basis},%
\footnote{
	See \labelcref{anchor:basis} for the definition of a valid operator basis that can be processed by \autoeft.
}
run the command:
\begin{Texinfoindented}
\begin{Texinfopreformatted}%
\ttfamily autoeft latex efts/sm-eft/6/basis
\end{Texinfopreformatted}
\end{Texinfoindented}
By default, \autoeft\ stores all \code{.tex} files under the directory
\file{tex/sm-eft/6/}.  If this directory does not contain a file called
\file{main.tex}, an appropriate file will be generated
automatically. From this \LaTeX\ file (for example with the
help of \code{pdflatex}), one can produce a \code{PDF} document which contains
a table encoding the model; a table representing the numbers of types, terms,
and operators per family; the respective Hilbert series; and the information
encoded in the operator files for each type.\footnote{Using the option
\code{-c}, one can directly compile \file{main.tex} with the \autoeft\ call,
see \labelcref{anchor:LaTeX}.} The \code{latex} command also supports the
\code{{-}{-}select} and \code{{-}{-}ignore} options, to restrict the
generation of \TeX\ files to a subset of the entire basis
(cf.~\cref{sec:constructing-operators} and \labelcref{anchor:LaTeX}).

%- }}}
%- }}}
%- {{{ section{Conclusion}

\section{Conclusion}\label{sec:conclusion}
We have presented \autoeft, a completely independent implementation of the
algorithms and concepts proposed in
\citeres{Fonseca:2019yya,Li:2020gnx,Li:2020xlh} for the automated bottom-up
construction of on-shell operator bases for generic \efts.  \autoeft\ has been
successfully used to construct---for the first time---the complete and
non-redundant \smeft\ and \grsmeft\ operator bases for mass dimensions 10, 11,
and 12~\cite{Harlander:2023psl}.  Besides the \sm, \autoeft\ can accommodate
various low-energy scenarios and generate the respective \eft\ operator bases,
in principle up to arbitrary mass dimension.  Due to the simple format of the
input files and the command-line utilities provided by \autoeft, the user can
compose custom models in a straight-forward way and construct \eft\ operator
bases with minimal effort. Its phenomenological purpose is to eliminate the
task of manually constructing \efts\ from low-energy theories which may
involve as-of-yet undiscovered light particles. But \autoeft\ may also help to
understand deeper structures of \efts\ at the general level (see,
e.g., \citere{Alminawi:2023qtf}).

\autoeft\ provides a foundation for future \eft\ frameworks and we plan to
extend its capabilities in various respects, including the automated
translation of the output to the
\code{FeynRules}~\cite{Christensen:2008py,Alloul:2013bka} format, or the
capability to transform Wilson coefficients between different bases. This
latter feature will also allow for a detailed comparison to programs of
similar purpose, which is currently impeded by different choices of the bases.

%- }}}
%- {{{ Acknowledgments:

\paragraph{Acknowledgments}
We would like to thank Jakob Linder and Maximilian Rzehak for helpful comments
and extensive $\beta$-tests, and Tim Kempkens for the inspiring collaboration
at the earlier stages of the project.  This research was supported by the
Deutsche Forschungsgemeinschaft (\abbrev{DFG}, German Research Foundation)
under grant 400140256~--~\textit{GRK~2497: The physics of the heaviest
  particles at the LHC}, and grant 396021762~--~\textit{TRR~257:
  P3H~--~Particle Physics Phenomenology after the Higgs Discovery}.

%- }}}
%- }}}
%- {{{ appendix:

\appendix

%- {{{ section{Relations to Conventional Notation}

\section{Relations to Conventional Notation}\label{sec:relations-to-conventional-notation}

%- {{{ intro:
In this section, we describe how the output of \autoeft\ can be related to a
more conventional notation, containing bispinors, field-strength tensors, and
Weyl tensors with Lorentz four-vector indices as well as fields carrying
anti-fundamental and adjoint indices of the internal symmetry groups.
Applying the following translation rules is straightforward but may result in large
expressions. Their systematic simplification requires a well-defined
criterion for \enquote{simplicity}, for example, and achieving maximal
simplicity may require computationally expensive permutation operations
which can exceed the capability of the available hardware.  We therefore
defer its implementation to future versions of \autoeft.

The four-component bispinors can be decomposed into left- and right-handed
part, such that they are equally represented by a single two component Weyl
spinor:
\begin{equation}
    \Psi_{\mathrm{L}} =
    \begin{pmatrix}
        {\psi_\mathrm{L}}_\alpha \\ 0
    \end{pmatrix}
    \,,\quad
    \bar\Psi_{\mathrm{L}} =
    \begin{pmatrix}
        0, {\psi_\mathrm{L}}^\dagger_{\dot\alpha}
    \end{pmatrix}
    \,,\quad
    \Psi_{\mathrm{R}} =
    \begin{pmatrix}
        0 \\ {\psi_\mathrm{R}}^{\dot\alpha}
    \end{pmatrix}
    \,,\quad
    \bar\Psi_{\mathrm{R}} =
    \begin{pmatrix}
        {\psi^\dagger_\mathrm{R}}^{\alpha}, 0
    \end{pmatrix}
	\,.
\end{equation}
Note that in the all-left chirality notation, $\psi\equiv\psi_{\mathrm{L}}$
and $\psi_{\mathbb{C}}\equiv\psi^\dagger_\mathrm{R}$.

The output of \autoeft\ containing Weyl spinors can be expressed in terms of bilinears involving the bispinors
\begin{equation}
    \Psi =
    \begin{pmatrix}
        {\psi_\mathrm{L}}_\alpha \\[.3em] {\psi_\mathrm{R}}^{\dot\alpha}
    \end{pmatrix}
    \,,\qquad
    \Chi =
    \begin{pmatrix}
        {\chi_\mathrm{L}}_\alpha \\[.3em] {\chi_\mathrm{R}}^{\dot\alpha}
    \end{pmatrix}
    \,,
\end{equation}
using the relations
\begin{equation}
\begin{aligned}
	\bar{\Psi}_{\mathrm{R}}\Chi_{\mathrm{L}} &= {\psi^\dagger_\mathrm{R}}^{\alpha} {\chi_\mathrm{L}}_\alpha
	\,,&
	\bar{\Psi}_{\mathrm{L}}\Chi_{\mathrm{R}} &= {\psi_\mathrm{L}}^\dagger_{\dot\alpha} {\chi_\mathrm{R}}^{\dot\alpha}
	\,,\\
	\bar{\Psi}_{\mathrm{R}}\gamma^\mu\Chi_{\mathrm{R}} &= {\psi^\dagger_\mathrm{R}}^{\alpha} {(\sigma^\mu)}_{\alpha\dot\alpha} {\chi_\mathrm{R}}^{\dot\alpha}
	\,,&
	\bar{\Psi}_{\mathrm{L}}\gamma^\mu\Chi_{\mathrm{L}} &= {\psi_\mathrm{L}}^\dagger_{\dot\alpha} {(\bar{\sigma}^\mu)}^{\dot\alpha\alpha} {\chi_\mathrm{L}}_\alpha
	\,,\\
	\bar{\Psi}_{\mathrm{R}} \sigma^{\mu\nu} \Chi_{\mathrm{L}} &= {\psi^\dagger_\mathrm{R}}^{\alpha} {(\sigma^{\mu\nu})}\indices{_\alpha^\beta} {\chi_\mathrm{L}}_\beta
	\,,&
	\bar{\Psi}_{\mathrm{L}}\sigma^{\mu\nu} \Chi_{\mathrm{R}} &= {\psi_\mathrm{L}}^\dagger_{\dot\alpha} {(\bar{\sigma}^{\mu\nu})}\indices{^{\dot\alpha}_{\dot\beta}} {\chi_\mathrm{R}}^{\dot\beta}
	\,,\\
	\Psi_{\mathrm{R}}^\T \cc \Chi_{\mathrm{R}} &= {\psi_\mathrm{R}}_{\dot\alpha} {\chi_\mathrm{R}}^{\dot\alpha}
	\,,&
	\Psi_{\mathrm{L}}^\T \cc \Chi_{\mathrm{L}} &= {\psi_\mathrm{L}}^\alpha {\psi_\mathrm{L}}_\alpha
	\,,\\
	\Psi_{\mathrm{R}}^\T \cc \gamma^\mu \Chi_{\mathrm{L}} &= {\psi_\mathrm{R}}_{\dot\alpha} {(\bar{\sigma}^\mu)}^{\dot\alpha\alpha} {\chi_\mathrm{L}}_\alpha
	\,,&
	\Psi_{\mathrm{L}}^\T \cc \gamma^\mu \Chi_{\mathrm{R}} &= {\psi_\mathrm{L}}^\alpha {(\sigma^\mu)}_{\alpha\dot\alpha} {\chi_\mathrm{R}}^{\dot\alpha}
	\,,\\
	\Psi_{\mathrm{R}}^\T \cc \sigma^{\mu\nu} \Chi_{\mathrm{R}} &= {\psi_\mathrm{R}}_{\dot\alpha} {(\bar{\sigma}^{\mu\nu})}\indices{^{\dot\alpha}_{\dot\beta}} {\chi_\mathrm{R}}^{\dot\beta}
	\,,&
	\Psi_{\mathrm{L}}^\T \cc \sigma^{\mu\nu} \Chi_{\mathrm{L}} &= {\psi_\mathrm{L}}^\alpha {(\sigma^{\mu\nu})}\indices{_\alpha^\beta} {\psi_\mathrm{L}}_\beta
	\,,\\
	\bar{\Psi}_{\mathrm{R}} \cc \bar{\Chi}_{\mathrm{R}}^\T &= {\psi^\dagger_\mathrm{R}}^{\alpha} {\chi^\dagger_\mathrm{R}}_{\alpha}
	\,,&
	\bar{\Psi}_{\mathrm{L}} \cc \bar{\Chi}_{\mathrm{L}}^\T &= {\psi_\mathrm{L}}^\dagger_{\dot\alpha} {\chi_\mathrm{L}^\dagger}^{\dot\alpha}
	\,,\\
	\bar{\Psi}_{\mathrm{R}} \gamma^\mu \cc \bar{\Chi}_{\mathrm{L}}^\T &= {\psi^\dagger_\mathrm{R}}^{\alpha} {(\sigma^\mu)}_{\alpha\dot\alpha} {\chi_\mathrm{L}^\dagger}^{\dot\alpha}
	\,,&
	\bar{\Psi}_{\mathrm{L}} \gamma^\mu \cc \bar{\Chi}_{\mathrm{R}}^\T &= {\psi_\mathrm{L}}^\dagger_{\dot\alpha} {(\bar{\sigma}^\mu)}^{\dot\alpha\alpha} {\chi^\dagger_\mathrm{R}}_{\alpha}
	\,,\\
	\bar{\Psi}_{\mathrm{R}} \sigma^{\mu\nu} \cc \bar{\Chi}_{\mathrm{R}}^\T &= {\psi^\dagger_\mathrm{R}}^{\alpha} {(\sigma^{\mu\nu})}\indices{_\alpha^\beta} {\chi^\dagger_\mathrm{R}}_{\beta}
	\,,&
	\bar{\Psi}_{\mathrm{L}} \sigma^{\mu\nu} \cc \bar{\Chi}_{\mathrm{L}}^\T &= {\psi_\mathrm{L}}^\dagger_{\dot\alpha} {(\bar{\sigma}^{\mu\nu})}\indices{^{\dot\alpha}_{\dot\beta}} {\chi_\mathrm{L}^\dagger}^{\dot\beta}
	\,,
\end{aligned}
\end{equation}
where
\begin{equation}
\begin{gathered}\label{eq:sigma}
    {(\sigma^\mu)}_{\alpha\dot\alpha} = {(I,\vec{\sigma})}_{\alpha\dot\alpha}
    \,,\qquad
    {(\bar\sigma^\mu)}^{\dot\alpha\alpha} = {(I,-\vec{\sigma})}^{\dot\alpha\alpha}
	\,,\\
    {(\sigma^{\mu\nu})}\indices{_\alpha^\beta} =
    \frac{i}{2}{\left(\sigma^\mu\bar\sigma^\nu-\sigma^\nu\bar\sigma^\mu\right)}\indices{_\alpha^\beta}
    \,,\qquad
    {(\bar{\sigma}_{\mu\nu})}\indices{^{\dot\alpha}_{\dot\beta}} =
    \frac{i}{2}{\left(\bar\sigma_\mu\sigma_\nu-\bar\sigma_\nu\sigma_\mu\right)}\indices{^{\dot\alpha}_{\dot\beta}}
    \,,
\end{gathered}
\end{equation}
and
\begin{equation}
\gamma^\mu =
\begin{pmatrix}
0 & {(\sigma^\mu)}_{\alpha\dot\beta} \\
{(\bar{\sigma}^\mu)}^{\dot\alpha\beta} & 0
\end{pmatrix}
\,,\qquad
\sigma^{\mu\nu} = \frac{i}{2}\left[\gamma^\mu,\gamma^\nu\right]
\,,\qquad
\cc = i\gamma^0\gamma^2
\,.
\end{equation}
Here, $I$ is the $2 \times 2$ identity matrix and $\vec{\sigma} =
\left(\sigma^1,\sigma^2,\sigma^3\right)$ denotes the Pauli matrices.

Using the normalization of Li~\textit{et al.}~\cite{Li:2023wdz}, the covariant
derivative, field-strength tensor, and Weyl tensor with Lorentz four-vector
indices are given by
\begin{equation}
\begin{gathered}
		D_\mu = -\frac12 D_{\alpha}^{\dot\alpha}{(\sigma_\mu)}^{\alpha}_{\dot\alpha}
        \,,\qquad
        D_\alpha^{\dot\alpha} = D_\mu {(\sigma^\mu)}_\alpha^{\dot\alpha}
		\,,
\end{gathered}
\end{equation}
\begin{equation}
\begin{gathered}
		F^{\mu\nu} =
		\frac{i}{4}
		F_{\mathrm{L}}^{\,\alpha\beta}
		\sigma^{\mu\nu}_{\alpha\beta}
		-
		\frac{i}{4}
		F_{\mathrm{R}}^{\,\dot\alpha\dot\beta}
		\bar{\sigma}^{\mu\nu}_{\dot\alpha\dot\beta}
		\,,\\
		F_{\mathrm{L}\,\alpha\beta} = \frac{i}{2}F_{\mu\nu}\sigma^{\mu\nu}_{\alpha\beta}
        \,,\qquad
        F_{\mathrm{R}}^{\,\dot\alpha\dot\beta} = -\frac{i}{2}F^{\mu\nu}\bar{\sigma}_{\mu\nu}^{\dot\alpha\dot\beta}
        \,,
\end{gathered}
\end{equation}
and
\begin{equation}
\begin{gathered}
		C^{\mu\nu\rho\sigma} =
    	-\frac{1}{16}C_\mathrm{L}^{\,\alpha\beta\gamma\delta}
		\sigma^{\mu\nu}_{\alpha\beta}
		\sigma^{\rho\sigma}_{\gamma\delta}
    	-\frac{1}{16}C_{\mathrm{R}}^{\,\dot\alpha\dot\beta\dot\gamma\dot\delta}
		\bar{\sigma}^{\mu\nu}_{\dot\alpha\dot\beta}
		\bar{\sigma}^{\rho\sigma}_{\dot\gamma\dot\delta}
		\,,\\
    	C_{\mathrm{L}\,\alpha\beta\gamma\delta} = -\frac{1}{4}C_{\mu\nu\rho\sigma}\sigma^{\mu\nu}_{\alpha\beta}\sigma^{\rho\sigma}_{\gamma\delta}
    	\,,\qquad
      	C_{\mathrm{R}}^{\,\dot\alpha\dot\beta\dot\gamma\dot\delta} = -\frac{1}{4}C^{\mu\nu\rho\sigma}\bar{\sigma}_{\mu\nu}^{\dot\alpha\dot\beta}\bar{\sigma}_{\rho\sigma}^{\dot\gamma\dot\delta}
      	\,,
\end{gathered}
\end{equation}
respectively.

Similarly, fields with $SU(N)$ anti-fundamental and adjoint indices are given by
\begin{gather}
	{(\psi^\dagger)}^b = \frac{1}{(N-1)!}\epsilon^{a_1\dots a_{N-1}b} \psi^\dagger_{a_1\dots a_{N-1}}
	\,,\qquad
	\psi^\dagger_{a_1\dots a_{N-1}} = \epsilon_{a_1\dots a_{N-1}b}{(\psi^\dagger)}^b \,,
\end{gather}
and
\begin{equation}
\begin{aligned}
	F^A = \frac{1}{T_F(N-1)!}\epsilon^{a_1a_3\dots a_{N-1}b}{(T^A)}\indices{^{a_2}_b} F_{a_1a_2a_3\dots a_N}
	\,,\\
	F_{a_1a_2a_3\dots a_N} = \epsilon_{a_1a_3\dots a_{N-1}b}{(T^A)}\indices{^b_{a_2}} F^A
	\,,
\end{aligned}
\end{equation}
respectively.  Here, $T^A$ denotes the $SU(N)$ generators in the fundamental
representation, and $T_F$ defines their normalization via $Tr(T^A T^B) =
T_F\,\delta^{AB}$.

%- }}}
%- }}}
%- {{{ section{appendix.tex}

\section{{Invoking \texttt{autoeft}}}
\label{anchor:Invoking}%

\index[cp]{autoeft synopsis@\texttt{autoeft \textnormal{synopsis}}}%
\subsection*{{Synopsis}}

\begin{Texinfoindented}
\begin{Texinfopreformatted}%
\ttfamily autoeft [\Texinfocommandstyletextvar{options}] \Texinfocommandstyletextvar{command}\ [\Texinfocommandstyletextvar{args}]
\end{Texinfopreformatted}
\end{Texinfoindented}

\subsection*{{Description}}

This is the main command starting \texttt{autoeft}.
Its general behavior is affected by the \Texinfocommandstyletextvar{options}.
\texttt{autoeft} offers several commands which are explained in the following sections.
The behavior of each \Texinfocommandstyletextvar{command} is controlled by \Texinfocommandstyletextvar{args}, including positional arguments and further options.

\subsection*{{Command-Line Options}}

\begin{description}
\item[{\parbox[b]{\linewidth}{%
\texttt{-h}\\
\texttt{{-}{-}help}}}]
\index[cp]{-h@\texttt{-h}}%
\index[cp]{--help@\texttt{{-}{-}help}}%
Print a usage message briefly summarizing the command-line options.

\item[{\parbox[b]{\linewidth}{%
\texttt{-v}\\
\texttt{{-}{-}version}}}]
\index[cp]{-v@\texttt{-v}}%
\index[cp]{--version@\texttt{{-}{-}version}}%
Print the version number of \texttt{autoeft} to the standard output stream.

\item[{\parbox[b]{\linewidth}{%
\texttt{-q}\\
\texttt{{-}{-}quiet}}}]
\index[cp]{-q@\texttt{-q}}%
\index[cp]{--quiet@\texttt{{-}{-}quiet}}%
Suppress all output to the standard output stream.
\end{description}

\subsection{{\texttt{autoeft} Commands}}
\label{anchor:Commands}%

\begin{description}
\item[{\parbox[b]{\linewidth}{%
\texttt{autoeft check}}}]
\index[cp]{check@\texttt{check}}%
Validate the installation of \texttt{autoeft} by comparing to a pre-determined basis.

\item[{\parbox[b]{\linewidth}{%
\texttt{autoeft sample-model}}}]
\index[cp]{sample-model@\texttt{sample-model}}%
Print the Standard Model definition in YAML format to the standard output stream.

\item[{\parbox[b]{\linewidth}{%
\texttt{autoeft construct}\\
\texttt{autoeft c}}}]
\index[cp]{construct@\texttt{construct}}%
\index[cp]{c@\texttt{c}}%
Construct an operator basis for a given model and mass dimension.

\item[{\parbox[b]{\linewidth}{%
\texttt{autoeft count}}}]
\index[cp]{count@\texttt{count}}%
Count the number of families, types, terms, and operators for a given basis.

\item[{\parbox[b]{\linewidth}{%
\texttt{autoeft latex}\\
\texttt{autoeft l}}}]
\index[cp]{latex@\texttt{latex}}%
\index[cp]{l@\texttt{l}}%
Generate and compile \TeX{} files for a given basis.

\item[{\parbox[b]{\linewidth}{%
\texttt{autoeft generators}\\
\texttt{autoeft g}}}]
\index[cp]{generators@\texttt{generators}}%
\index[cp]{g@\texttt{g}}%
View or create the symmetric group representation generators.
\end{description}

\subsubsection{{\texttt{check} Command}}
\label{anchor:Check}%

\index[cp]{check synopsis@\texttt{check \textnormal{synopsis}}}%
\subsubsection*{{Synopsis}}

\begin{Texinfoindented}
\begin{Texinfopreformatted}%
\ttfamily autoeft check
\end{Texinfopreformatted}
\end{Texinfoindented}

\subsubsection*{{Description}}
This command constructs the mass dimension-six operator basis for the Standard Model Effective Field Theory and compares it to a pre-determined basis.
If both bases agree, the installation is reported as successful.

\subsubsection{{\texttt{sample-model} Command}}
\label{anchor:Sample}%

\index[cp]{sample-model synopsis@\texttt{sample-model \textnormal{synopsis}}}%
\subsubsection*{{Synopsis}}

\begin{Texinfoindented}
\begin{Texinfopreformatted}%
\ttfamily autoeft sample-model
\end{Texinfopreformatted}
\end{Texinfoindented}

\subsubsection*{{Description}}
This command prints the content of the model file (see \hyperref[anchor:Model]{[Model], page~\pageref*{anchor:Model}}) for the Standard Model to the standard output stream.
To obtain the model file \texttt{sm.yml} in the current working directory, simply run:
\begin{Texinfoindented}
\begin{Texinfopreformatted}%
\ttfamily autoeft sample-model > sm.yml
\end{Texinfopreformatted}
\end{Texinfoindented}

\subsubsection{{\texttt{construct} Command}}
\label{anchor:Construct}%

\index[cp]{construct synopsis@\texttt{construct \textnormal{synopsis}}}%
\subsubsection*{{Synopsis}}

\begin{Texinfoindented}
\begin{Texinfopreformatted}%
\ttfamily autoeft construct [\Texinfocommandstyletextvar{options}] \Texinfocommandstyletextvar{model}\ \Texinfocommandstyletextvar{dimension}
\end{Texinfopreformatted}
\end{Texinfoindented}

\subsubsection*{{Description}}
This command starts the construction of an operator basis.
The details of the basis depend on the provided model file \texttt{model.yml} (see \hyperref[anchor:Model]{[Model], page~\pageref*{anchor:Model}}) and mass dimension \texttt{D}.
The result will be saved in the (default) output directory under \texttt{<model>-eft/<D>/basis/}.
The operators are further collected by their \texttt{family} (see \hyperref[anchor:family]{[\textsl{family}], page~\pageref*{anchor:family}}) and \texttt{type} (see \hyperref[anchor:type]{[\textsl{type}], page~\pageref*{anchor:type}}) in subdirectories of structure \texttt{<N>/<family>/<type>.yml}, where \texttt{N} is the number of fields in the operator.
See \hyperref[anchor:Output]{[Output], page~\pageref*{anchor:Output}} for the format of the output.

\subsubsection*{{Positional Arguments}}

\begin{description}
\item[{\parbox[b]{\linewidth}{%
\Texinfocommandstyletextvar{model}}}]
Path to the model file that should be used for the construction (see \hyperref[anchor:Model]{[Model], page~\pageref*{anchor:Model}}).

\item[{\parbox[b]{\linewidth}{%
\Texinfocommandstyletextvar{dimension}}}]
The integer mass dimension the constructed operators should have.
\end{description}

\subsubsection*{{Options}}

\begin{description}
\item[{\parbox[b]{\linewidth}{%
\texttt{-h}\\
\texttt{{-}{-}help}}}]
\index[cp]{-h (construct)@\texttt{-h \textnormal{(construct)}}}%
\index[cp]{--help (construct)@\texttt{{-}{-}help \textnormal{(construct)}}}%
Print a usage message briefly summarizing the command-line options.

\item[{\parbox[b]{\linewidth}{%
\texttt{-v}\\
\texttt{{-}{-}verbose}}}]
\index[cp]{-v (construct)@\texttt{-v \textnormal{(construct)}}}%
\index[cp]{--verbose (construct)@\texttt{{-}{-}verbose \textnormal{(construct)}}}%
Print a tree-like structure of operator families and types during the construction.

\item[{\parbox[b]{\linewidth}{%
\texttt{-t \Texinfocommandstyletextvar{n}}\\
\texttt{{-}{-}threads=\Texinfocommandstyletextvar{n}}}}]
\index[cp]{-t (construct)@\texttt{-t \textnormal{(construct)}}}%
\index[cp]{--threads (construct)@\texttt{{-}{-}threads \textnormal{(construct)}}}%
Set the number of threads \Texinfocommandstyletextvar{n} that start \texttt{form} processes.
By default, \Texinfocommandstyletextvar{n} is the number of CPUs in the system.

\item[{\parbox[b]{\linewidth}{%
\texttt{-n \Texinfocommandstyletextvar{name}}\\
\texttt{{-}{-}name=\Texinfocommandstyletextvar{name}}}}]
\index[cp]{-n (construct)@\texttt{-n \textnormal{(construct)}}}%
\index[cp]{--name (construct)@\texttt{{-}{-}name \textnormal{(construct)}}}%
Set the EFT \Texinfocommandstyletextvar{name}.
By default, the name is \texttt{<model>-eft}.

\item[{\parbox[b]{\linewidth}{%
\texttt{-o \Texinfocommandstyletextvar{path}}\\
\texttt{{-}{-}output=\Texinfocommandstyletextvar{path}}}}]
\index[cp]{-o (construct)@\texttt{-o \textnormal{(construct)}}}%
\index[cp]{--output (construct)@\texttt{{-}{-}output \textnormal{(construct)}}}%
Set the output \Texinfocommandstyletextvar{path}, where the operators will be saved.
By default, the operators are saved under \texttt{efts/} in the current working directory.

\item[{\parbox[b]{\linewidth}{%
\texttt{-s \Texinfocommandstyletextvar{pattern}}\\
\texttt{{-}{-}select=\Texinfocommandstyletextvar{pattern}}}}]
\index[cp]{-s (construct)@\texttt{-s \textnormal{(construct)}}}%
\index[cp]{--select (construct)@\texttt{{-}{-}select \textnormal{(construct)}}}%
Only construct operator types (see \hyperref[anchor:type]{[\textsl{type}], page~\pageref*{anchor:type}}) that match with \Texinfocommandstyletextvar{pattern}.
The pattern must be given as a string representing a mapping---denoted by curly braces---from the fields to their number of occurrences in the desired type (e.g., `\texttt{\{H:\ 3,\ H+:\ 3\}}').
Besides explicit numbers, the symbol `\texttt{+}' can be used to require at least one occurrence.
Alternatively, a range can be provided by `\texttt{x..y}', meaning there must be at least `\texttt{x}' and at most `\texttt{y}' occurrences.
If one of the bounds is omitted (e.g., `\texttt{x..}', or `\texttt{..y}'), only the remaining one is enforced.
By default, no selection is performed and any operator will be constructed.
If this option is used multiple times, an operator type must match at least one pattern to be constructed.
If this option is combined with the \texttt{-i}(\texttt{{-}{-}ignore}) option, \emph{select} is applied before \emph{ignore}.

\item[{\parbox[b]{\linewidth}{%
\texttt{-i \Texinfocommandstyletextvar{pattern}}\\
\texttt{{-}{-}ignore=\Texinfocommandstyletextvar{pattern}}}}]
\index[cp]{-i (construct)@\texttt{-i \textnormal{(construct)}}}%
\index[cp]{--ignore (construct)@\texttt{{-}{-}ignore \textnormal{(construct)}}}%
Do not construct operator types (see \hyperref[anchor:type]{[\textsl{type}], page~\pageref*{anchor:type}}) that match with \Texinfocommandstyletextvar{pattern}.
The pattern must be given as a string representing a mapping---denoted by curly braces---from the fields to their number of occurrences in the desired type (e.g., `\texttt{\{H:\ 3,\ H+:\ 3\}}').
Besides explicit numbers, the symbol `\texttt{+}' can be used to require at least one occurrence.
Alternatively, a range can be provided by `\texttt{x..y}', meaning there must be at least `\texttt{x}' and at most `\texttt{y}' occurrences.
If one of the bounds is omitted (e.g., `\texttt{x..}', or `\texttt{..y}'), only the remaining one is enforced.
By default, no exclusion is performed and all operators will be constructed.
If this option is used multiple times, an operator type that matches at least one pattern will not be constructed.
If this option is combined with the \texttt{-s}(\texttt{{-}{-}select}) option, \emph{select} is applied before \emph{ignore}.

\item[{\parbox[b]{\linewidth}{%
\texttt{{-}{-}dry-run}}}]
\index[cp]{--dry-run (construct)@\texttt{{-}{-}dry-run \textnormal{(construct)}}}%
Only list the operator types (see \hyperref[anchor:type]{[\textsl{type}], page~\pageref*{anchor:type}}) that match the provided selection.
No explicit construction is performed.

\item[{\parbox[b]{\linewidth}{%
\texttt{{-}{-}generators=\Texinfocommandstyletextvar{path}}}}]
\index[cp]{--generators (construct)@\texttt{{-}{-}generators \textnormal{(construct)}}}%
Set the generators \Texinfocommandstyletextvar{path}, where \texttt{autoeft} searches for the symmetric group representation generators (see \hyperref[anchor:Generators]{[Generators], page~\pageref*{anchor:Generators}}).
By default, this is set to \texttt{gens/} in the current working directory.
If the directory does not exist or some generator files are missing, \texttt{autoeft} loads fallback generators for the representations up to S\textsubscript{9}.

\item[{\parbox[b]{\linewidth}{%
\texttt{{-}{-}overwrite}}}]
\index[cp]{--overwrite (construct)@\texttt{{-}{-}overwrite \textnormal{(construct)}}}%
Overwrite existing operator files in the output directory.

\item[{\parbox[b]{\linewidth}{%
\texttt{{-}{-}no\_hc}}}]
\index[cp]{--no\_hc (construct)@\texttt{{-}{-}no\_hc \textnormal{(construct)}}}%
Prevent the explicit construction of conjugate operator types.
\end{description}

\subsubsection{{\texttt{count} Command}}
\label{anchor:Count}%

\index[cp]{count synopsis@\texttt{count \textnormal{synopsis}}}%
\subsubsection*{{Synopsis}}

\begin{Texinfoindented}
\begin{Texinfopreformatted}%
\ttfamily autoeft count [\Texinfocommandstyletextvar{options}] \Texinfocommandstyletextvar{basis}
\end{Texinfopreformatted}
\end{Texinfoindented}

\subsubsection*{{Description}}
This command counts the number of families, types, terms, and operators for a given basis (see \hyperref[anchor:Vocabulary]{[Vocabulary], page~\pageref*{anchor:Vocabulary}}).

\subsubsection*{{Positional Arguments}}

\begin{description}
\item[{\parbox[b]{\linewidth}{%
\Texinfocommandstyletextvar{basis}}}]
Path to the basis containing the operators (see \hyperref[anchor:basis]{[\textsl{basis}], page~\pageref*{anchor:basis}}).
\end{description}

\subsubsection*{{Options}}

\begin{description}
\item[{\parbox[b]{\linewidth}{%
\texttt{-h}\\
\texttt{{-}{-}help}}}]
\index[cp]{-h (count)@\texttt{-h \textnormal{(count)}}}%
\index[cp]{--help (count)@\texttt{{-}{-}help \textnormal{(count)}}}%
Print a usage message briefly summarizing the command-line options.

\item[{\parbox[b]{\linewidth}{%
\texttt{-v}\\
\texttt{{-}{-}verbose}}}]
\index[cp]{-v (count)@\texttt{-v \textnormal{(count)}}}%
\index[cp]{--verbose (count)@\texttt{{-}{-}verbose \textnormal{(count)}}}%
Print a tree-like structure of operator families and types.

\item[{\parbox[b]{\linewidth}{%
\texttt{-o \Texinfocommandstyletextvar{file}}\\
\texttt{{-}{-}output=\Texinfocommandstyletextvar{file}}}}]
\index[cp]{-o (count)@\texttt{-o \textnormal{(count)}}}%
\index[cp]{--output (count)@\texttt{{-}{-}output \textnormal{(count)}}}%
Set the output \Texinfocommandstyletextvar{file}, where the numbers will be saved.
By default, the numbers are saved in the file \texttt{counts.yml} in the current working directory.

\item[{\parbox[b]{\linewidth}{%
\texttt{-s \Texinfocommandstyletextvar{pattern}}\\
\texttt{{-}{-}select=\Texinfocommandstyletextvar{pattern}}}}]
\index[cp]{-s (count)@\texttt{-s \textnormal{(count)}}}%
\index[cp]{--select (count)@\texttt{{-}{-}select \textnormal{(count)}}}%
Only count operator types (see \hyperref[anchor:type]{[\textsl{type}], page~\pageref*{anchor:type}}) that match with \Texinfocommandstyletextvar{pattern}.
The pattern must be given as a string representing a mapping---denoted by curly braces---from the fields to their number of occurrences in the desired type (e.g., `\texttt{\{H:\ 3,\ H+:\ 3\}}').
Besides explicit numbers, the symbol `\texttt{+}' can be used to require at least one occurrence.
Alternatively, a range can be provided by `\texttt{x..y}', meaning there must be at least `\texttt{x}' and at most `\texttt{y}' occurrences.
If one of the bounds is omitted (e.g., `\texttt{x..}', or `\texttt{..y}'), only the remaining one is enforced.
By default, no selection is performed and any operator will be counted.
If this option is used multiple times, an operator type must match at least one pattern to be counted.
If this option is combined with the \texttt{-i}(\texttt{{-}{-}ignore}) option, \emph{select} is applied before \emph{ignore}.

\item[{\parbox[b]{\linewidth}{%
\texttt{-i \Texinfocommandstyletextvar{pattern}}\\
\texttt{{-}{-}ignore=\Texinfocommandstyletextvar{pattern}}}}]
\index[cp]{-i (count)@\texttt{-i \textnormal{(count)}}}%
\index[cp]{--ignore (count)@\texttt{{-}{-}ignore \textnormal{(count)}}}%
Do not count operator types (see \hyperref[anchor:type]{[\textsl{type}], page~\pageref*{anchor:type}}) that match with \Texinfocommandstyletextvar{pattern}.
The pattern must be given as a string representing a mapping---denoted by curly braces---from the fields to their number of occurrences in the desired type (e.g., `\texttt{\{H:\ 3,\ H+:\ 3\}}').
Besides explicit numbers, the symbol `\texttt{+}' can be used to require at least one occurrence.
Alternatively, a range can be provided by `\texttt{x..y}', meaning there must be at least `\texttt{x}' and at most `\texttt{y}' occurrences.
If one of the bounds is omitted (e.g., `\texttt{x..}', or `\texttt{..y}'), only the remaining one is enforced.
By default, no exclusion is performed and all operators will be counted.
If this option is used multiple times, an operator type that matches at least one pattern will not be counted.
If this option is combined with the \texttt{-s}(\texttt{{-}{-}select}) option, \emph{select} is applied before \emph{ignore}.

\item[{\parbox[b]{\linewidth}{%
\texttt{{-}{-}dry-run}}}]
\index[cp]{--dry-run (count)@\texttt{{-}{-}dry-run \textnormal{(count)}}}%
Only list the operator types (see \hyperref[anchor:type]{[\textsl{type}], page~\pageref*{anchor:type}}) that match the provided selection.
No explicit counting is performed.

\item[{\parbox[b]{\linewidth}{%
\texttt{{-}{-}no\_hc}}}]
\index[cp]{--no\_hc (count)@\texttt{{-}{-}no\_hc \textnormal{(count)}}}%
Prevent the implicit counting of conjugate operator types.
\end{description}

\subsubsection{{\texttt{latex} Command}}
\label{anchor:LaTeX}%

\index[cp]{latex synopsis@\texttt{latex \textnormal{synopsis}}}%
\subsubsection*{{Synopsis}}

\begin{Texinfoindented}
\begin{Texinfopreformatted}%
\ttfamily autoeft latex [\Texinfocommandstyletextvar{options}] \Texinfocommandstyletextvar{basis}
\end{Texinfopreformatted}
\end{Texinfoindented}

\subsubsection*{{Description}}
This command generates \TeX{} files for a given basis (see \hyperref[anchor:basis]{[\textsl{basis}], page~\pageref*{anchor:basis}}). The \TeX{} files represent all the information encoded in the operator files as \LaTeX{} markup.
The resulting files compose a valid \LaTeX{} document that can be compiled to a single PDF file.
\subsubsection*{{Positional Arguments}}

\begin{description}
\item[{\parbox[b]{\linewidth}{%
\Texinfocommandstyletextvar{basis}}}]
Path to the basis containing the operators (see \hyperref[anchor:basis]{[\textsl{basis}], page~\pageref*{anchor:basis}}).
\end{description}

\subsubsection*{{Options}}

\begin{description}
\item[{\parbox[b]{\linewidth}{%
\texttt{-h}\\
\texttt{{-}{-}help}}}]
\index[cp]{-h (latex)@\texttt{-h \textnormal{(latex)}}}%
\index[cp]{--help (latex)@\texttt{{-}{-}help \textnormal{(latex)}}}%
Print a usage message briefly summarizing the command-line options.

\item[{\parbox[b]{\linewidth}{%
\texttt{-c \Texinfocommandstyletextvar{command}}\\
\texttt{{-}{-}compile=\Texinfocommandstyletextvar{command}}}}]
\index[cp]{-c (latex)@\texttt{-c \textnormal{(latex)}}}%
\index[cp]{--compile (latex)@\texttt{{-}{-}compile \textnormal{(latex)}}}%
Compile the \TeX{} files by invoking the \Texinfocommandstyletextvar{command} in the output directory.

\item[{\parbox[b]{\linewidth}{%
\texttt{-o \Texinfocommandstyletextvar{path}}\\
\texttt{{-}{-}output=\Texinfocommandstyletextvar{path}}}}]
\index[cp]{-o (latex)@\texttt{-o \textnormal{(latex)}}}%
\index[cp]{--output (latex)@\texttt{{-}{-}output \textnormal{(latex)}}}%
Set the output \Texinfocommandstyletextvar{path}, where the \TeX{} files will be saved.
By default, the \TeX{} files are saved under \texttt{tex/} in the current working directory.

\item[{\parbox[b]{\linewidth}{%
\texttt{-s \Texinfocommandstyletextvar{pattern}}\\
\texttt{{-}{-}select=\Texinfocommandstyletextvar{pattern}}}}]
\index[cp]{-s (latex)@\texttt{-s \textnormal{(latex)}}}%
\index[cp]{--select (latex)@\texttt{{-}{-}select \textnormal{(latex)}}}%
Only generate \TeX{} files for operator types (see \hyperref[anchor:type]{[\textsl{type}], page~\pageref*{anchor:type}}) that match with \Texinfocommandstyletextvar{pattern}.
The pattern must be given as a string representing a mapping---denoted by curly braces---from the fields to their number of occurrences in the desired type (e.g., `\texttt{\{H:\ 3,\ H+:\ 3\}}').
Besides explicit numbers, the symbol `\texttt{+}' can be used to require at least one occurrence.
Alternatively, a range can be provided by `\texttt{x..y}', meaning there must be at least `\texttt{x}' and at most `\texttt{y}' occurrences.
If one of the bounds is omitted (e.g., `\texttt{x..}', or `\texttt{..y}'), only the remaining one is enforced.
By default, no selection is performed and any operator will be included.
If this option is used multiple times, an operator type must match at least one pattern to be included.
If this option is combined with the \texttt{-i}(\texttt{{-}{-}ignore}) option, \emph{select} is applied before \emph{ignore}.

\item[{\parbox[b]{\linewidth}{%
\texttt{-i \Texinfocommandstyletextvar{pattern}}\\
\texttt{{-}{-}ignore=\Texinfocommandstyletextvar{pattern}}}}]
\index[cp]{-i (latex)@\texttt{-i \textnormal{(latex)}}}%
\index[cp]{--ignore (latex)@\texttt{{-}{-}ignore \textnormal{(latex)}}}%
Do not generate \TeX{} files for operator types (see \hyperref[anchor:type]{[\textsl{type}], page~\pageref*{anchor:type}}) that match with \Texinfocommandstyletextvar{pattern}.
The pattern must be given as a string representing a mapping---denoted by curly braces---from the fields to their number of occurrences in the desired type (e.g., `\texttt{\{H:\ 3,\ H+:\ 3\}}').
Besides explicit numbers, the symbol `\texttt{+}' can be used to require at least one occurrence.
Alternatively, a range can be provided by `\texttt{x..y}', meaning there must be at least `\texttt{x}' and at most `\texttt{y}' occurrences.
If one of the bounds is omitted (e.g., `\texttt{x..}', or `\texttt{..y}'), only the remaining one is enforced.
By default, no exclusion is performed and all operators will be included.
If this option is used multiple times, an operator type that matches at least one pattern will be excluded.
If this option is combined with the \texttt{-s}(\texttt{{-}{-}select}) option, \emph{select} is applied before \emph{ignore}.

\item[{\parbox[b]{\linewidth}{%
\texttt{{-}{-}dry-run}}}]
\index[cp]{--dry-run (latex)@\texttt{{-}{-}dry-run \textnormal{(latex)}}}%
Only list the operator types (see \hyperref[anchor:type]{[\textsl{type}], page~\pageref*{anchor:type}}) that match the provided selection.
No explicit \TeX{} files are generated.
\end{description}

\subsubsection{{\texttt{generators} Command}}
\label{anchor:Generators}%

\index[cp]{generators synopsis@\texttt{generators \textnormal{synopsis}}}%
\subsubsection*{{Synopsis}}

\begin{Texinfoindented}
\begin{Texinfopreformatted}%
\ttfamily autoeft generators [\Texinfocommandstyletextvar{options}]
\end{Texinfopreformatted}
\end{Texinfoindented}

\subsubsection*{{Description}}
This command handles the pre-computed generator matrices for the symmetric group representations.
If this command is executed \emph{without} the \texttt{-S} or \texttt{-P} options, a table of all generators that \texttt{autoeft} would load with the current options is printed to the standard output stream.
Otherwise, the respective generators are computed and stored.

\subsubsection*{{Options}}

\begin{description}
\item[{\parbox[b]{\linewidth}{%
\texttt{-h}\\
\texttt{{-}{-}help}}}]
\index[cp]{-h (generators)@\texttt{-h \textnormal{(generators)}}}%
\index[cp]{--help (generators)@\texttt{{-}{-}help \textnormal{(generators)}}}%
Print a usage message briefly summarizing the command-line options.

\item[{\parbox[b]{\linewidth}{%
\texttt{-o \Texinfocommandstyletextvar{path}}\\
\texttt{{-}{-}output=\Texinfocommandstyletextvar{path}}}}]
\index[cp]{-o (generators)@\texttt{-o \textnormal{(generators)}}}%
\index[cp]{--output (generators)@\texttt{{-}{-}output \textnormal{(generators)}}}%
Set the output \Texinfocommandstyletextvar{path}, where the generators will be saved.
By default, the generators are saved under \texttt{gens/} in the current working directory.

\item[{\parbox[b]{\linewidth}{%
\texttt{-S \Texinfocommandstyletextvar{N}}}}]
\index[cp]{-S (generators)@\texttt{-S \textnormal{(generators)}}}%
Create the generators for all irreducible representations of the symmetric group S\textsubscript{\Texinfocommandstyletextvar{N}} of degree \Texinfocommandstyletextvar{N}.

\item[{\parbox[b]{\linewidth}{%
\texttt{-P \Texinfocommandstyletextvar{p}\ [\Texinfocommandstyletextvar{p}\ \dots{}\@]}}}]
\index[cp]{-P (generators)@\texttt{-P \textnormal{(generators)}}}%
Create the generators for the irreducible representation given by the partition as a non-increasing list of integers \Texinfocommandstyletextvar{p}.

\item[{\parbox[b]{\linewidth}{%
\texttt{{-}{-}overwrite}}}]
\index[cp]{--overwrite (generators)@\texttt{{-}{-}overwrite \textnormal{(generators)}}}%
Overwrite existing generator files in the output directory.
\end{description}

\subsection{{Environment Variables}}
\label{anchor:Environment}%

\begin{description}
\item[{\parbox[b]{\linewidth}{%
\texttt{AUTOEFT\_PATH}}}]
\index[cp]{AUTOEFT\_PATH (environment variable)@\texttt{AUTOEFT\_PATH \textnormal{(environment variable)}}}%
The environment variable \texttt{AUTOEFT\_PATH} can be set to a path (or a list of paths, separated by `\texttt{:}'), to specify where \texttt{autoeft} searches for the \texttt{form} executable.
If \texttt{AUTOEFT\_PATH} is not set, the system \texttt{PATH} will be used instead.

\item[{\parbox[b]{\linewidth}{%
\texttt{AUTOEFT\_CS}}}]
\index[cp]{AUTOEFT\_CS (environment variable)@\texttt{AUTOEFT\_CS \textnormal{(environment variable)}}}%
The symbol appended to the field name to denote conjugate fields.
By default, the symbol `\texttt{+}' is used.

\item[{\parbox[b]{\linewidth}{%
\texttt{AUTOEFT\_DS}}}]
\index[cp]{AUTOEFT\_DS (environment variable)@\texttt{AUTOEFT\_DS \textnormal{(environment variable)}}}%
The symbol appended to spinor indices to denote dotted indices.
By default, the symbol `\texttt{\~{}}' is used.
\end{description}

\section{{Model File}}
\label{anchor:Model}%

\begin{description}
\item[{\parbox[b]{\linewidth}{%
\texttt{name \Texinfocommandstyletextvar{Scalar}}}}]
\index[cp]{name (model)@\texttt{name \textnormal{(model)}}}%
The name of the model.
This is usually a short identifier like `\texttt{SMEFT}' or `\texttt{LEFT}'.

\item[{\parbox[b]{\linewidth}{%
\texttt{description \Texinfocommandstyletextvar{Scalar}}}}]
\index[cp]{description (model)@\texttt{description \textnormal{(model)}}}%
Optional (longer) description of the model.

\item[{\parbox[b]{\linewidth}{%
\texttt{symmetries \Texinfocommandstyletextvar{Mapping}}}}]
\index[cp]{symmetries (model)@\texttt{symmetries \textnormal{(model)}}}%
Definition of the model's symmetries.
The symmetries are divided into the sub-entries \texttt{lorentz\_group}, \texttt{sun\_groups}, and \texttt{u1\_groups}.

\begin{description}
\item[{\parbox[b]{\linewidth}{%
\texttt{lorentz\_group \Texinfocommandstyletextvar{Mapping}}}}]
\index[cp]{lorentz\_group (symmetry)@\texttt{lorentz\_group \textnormal{(symmetry)}}}%
Properties of the Lorentz group---realized as $SU(2)_l \times SU(2)_r$.
If omitted, \texttt{autoeft} loads the Lorentz group with default values.

\begin{description}
\item[{\parbox[b]{\linewidth}{%
\texttt{name \Texinfocommandstyletextvar{Scalar}}}}]
\index[cp]{name (Lorentz group)@\texttt{name \textnormal{(Lorentz group)}}}%
The name associated with the Lorentz group.
\begin{itemize}[label=\textbullet{}]
\item Group names must start with a letter (`\texttt{A-z}').
\item Group names can only contain alpha-numeric characters and parentheses (`\texttt{A-z}', `\texttt{0-9}', `\texttt{(}', and `\texttt{)}').
\item Group names must end with an alpha-numeric character or a parenthesis (`\texttt{A-z}', `\texttt{0-9}', `\texttt{(}', or `\texttt{)}').
\end{itemize}
By default, the name `\texttt{Lorentz}' is used.

\item[{\parbox[b]{\linewidth}{%
\texttt{tex \Texinfocommandstyletextvar{Scalar}}}}]
\index[cp]{tex (Lorentz group)@\texttt{tex \textnormal{(Lorentz group)}}}%
The \TeX{} string associated with the Lorentz group.
By default, the group name surrounded by \texttt{\textbackslash{}mathtt\{\textbullet{}\}} is used.

\item[{\parbox[b]{\linewidth}{%
\texttt{indices \Texinfocommandstyletextvar{Sequence}}}}]
\index[cp]{indices (Lorentz group)@\texttt{indices \textnormal{(Lorentz group)}}}%
The list of \TeX{} (spinor) indices associated with the Lorentz group.
By default, the Greek letters `\texttt{$\alpha,\beta,\dots,\lambda$}' are used.
\end{description}

\item[{\parbox[b]{\linewidth}{%
\texttt{sun\_groups \Texinfocommandstyletextvar{Mapping}}}}]
\index[cp]{sun\_groups (symmetry)@\texttt{sun\_groups \textnormal{(symmetry)}}}%
Definition of the model's non-abelian symmetries---realized as $SU(N)$ groups.
Each entry is a mapping from the symmetry's name (e.g.,~`\texttt{QCD}' or `\texttt{SU(3)}') to its properties.
\begin{itemize}[label=\textbullet{}]
\item Group names must start with a letter (`\texttt{A-z}').
\item Group names can only contain alpha-numeric characters and parentheses (`\texttt{A-z}', `\texttt{0-9}', `\texttt{(}', and `\texttt{)}').
\item Group names must end with an alpha-numeric character or a parenthesis (`\texttt{A-z}', `\texttt{0-9}', `\texttt{(}', or `\texttt{)}').
\end{itemize}
By default, \texttt{sun\_groups} is the empty mapping `\texttt{\{\}}'.

\begin{description}
\item[{\parbox[b]{\linewidth}{%
\texttt{N \Texinfocommandstyletextvar{Scalar}}}}]
\index[cp]{N (SU(N) group)@\texttt{N \textnormal{($SU(N)$ group)}}}%
The degree of the respective $SU(N)$ group.
This must be a positive integer greater than `\texttt{1}'.

\item[{\parbox[b]{\linewidth}{%
\texttt{tex \Texinfocommandstyletextvar{Scalar}}}}]
\index[cp]{tex (SU(N) group)@\texttt{tex \textnormal{($SU(N)$ group)}}}%
The \TeX{} string associated with the respective $SU(N)$ group.
By default, the group name surrounded by \texttt{\textbackslash{}mathtt\{\textbullet{}\}} is used.

\item[{\parbox[b]{\linewidth}{%
\texttt{indices \Texinfocommandstyletextvar{Sequence}}}}]
\index[cp]{indices (SU(N) group)@\texttt{indices \textnormal{($SU(N)$ group)}}}%
The list of \TeX{} (fundamental) indices associated with the respective $SU(N)$ group.
By default, the characters `\texttt{a,b,\dots{}\@,k}' are used.
\end{description}

\item[{\parbox[b]{\linewidth}{%
\texttt{u1\_groups \Texinfocommandstyletextvar{Mapping}}}}]
\index[cp]{u1\_groups (symmetry)@\texttt{u1\_groups \textnormal{(symmetry)}}}%
Definition of the model's abelian symmetries---realized as $U(1)$ groups.
Each entry is a mapping from the symmetry's name (e.g.,~`\texttt{QED}' or `\texttt{U(1)}') to its properties.
\begin{itemize}[label=\textbullet{}]
\item Group names must start with a letter (`\texttt{A-z}').
\item Group names can only contain alpha-numeric characters and parentheses (`\texttt{A-z}', `\texttt{0-9}', `\texttt{(}', and `\texttt{)}').
\item Group names must end with an alpha-numeric character or a parenthesis (`\texttt{A-z}', `\texttt{0-9}', `\texttt{(}', or `\texttt{)}').
\end{itemize}
By default, \texttt{u1\_groups} is the empty mapping `\texttt{\{\}}'.

\begin{description}
\item[{\parbox[b]{\linewidth}{%
\texttt{violation \Texinfocommandstyletextvar{Scalar}}}}]
\index[cp]{violation (U(1) group)@\texttt{violation \textnormal{($U(1)$ group)}}}%
A unit of charge the operators are \textbf{allowed} to deviate from the exact $U(1)$ symmetry.
The unit can be an integer value (e.g.,~`\texttt{1}') or a fractional value (e.g.,~`\texttt{1/2}').
The \texttt{violation} $v$ can be combined with a \texttt{residual} charge $R$, such that only operators with total charge $Q$ satisfying $|Q - R| \le v$ are allowed.
The default value for \texttt{violation} is `\texttt{0}'.

\item[{\parbox[b]{\linewidth}{%
\texttt{residual \Texinfocommandstyletextvar{Scalar}}}}]
\index[cp]{residual (U(1) group)@\texttt{residual \textnormal{($U(1)$ group)}}}%
A unit of charge the operators \textbf{must} deviate from the exact $U(1)$ symmetry.
The unit can be an integer value (e.g.,~`\texttt{1}') or a fractional value (e.g.,~`\texttt{1/2}'). See also \texttt{violation}.
The default value for \texttt{residual} is `\texttt{0}'.

\item[{\parbox[b]{\linewidth}{%
\texttt{tex \Texinfocommandstyletextvar{Scalar}}}}]
\index[cp]{tex (U(1) group)@\texttt{tex \textnormal{($U(1)$ group)}}}%
The \TeX{} string associated with the respective $U(1)$ group.
By default, the group name surrounded by \texttt{\textbackslash{}mathtt\{\textbullet{}\}} is used.
\end{description}
\end{description}

\item[{\parbox[b]{\linewidth}{%
\texttt{fields \Texinfocommandstyletextvar{Mapping}}}}]
\index[cp]{fields (model)@\texttt{fields \textnormal{(model)}}}%
Definition of the model's field content.
Each entry is a mapping from the field's name (e.g.,~`\texttt{Q}' or `\texttt{H}') to its properties.
\begin{itemize}[label=\textbullet{}]
\item Field names must start with a letter (`\texttt{A-z}').
\item Field names can only contain alpha-numeric characters (`\texttt{A-z}' and `\texttt{0-9}').
\item Field names must end with an alpha-numeric character or a plus (`\texttt{A-z}', `\texttt{0-9}', or `\texttt{+}').
If the environment variable \texttt{AUTOEFT\_CS} is set, it replaces the plus symbol `\texttt{+}' in the above restrictions (see \hyperref[anchor:Environment]{[Environment], page~\pageref*{anchor:Environment}}).
\item The field name cannot be `\texttt{D}', as this symbol is reserved for the covariant derivative .
\end{itemize}
By default, \texttt{fields} is the empty mapping `\texttt{\{\}}'.

\begin{description}
\item[{\parbox[b]{\linewidth}{%
\texttt{representations \Texinfocommandstyletextvar{Mapping}}}}]
\index[cp]{representations (field)@\texttt{representations \textnormal{(field)}}}%
Definition of the field's representations.
Each entry is a mapping from a group name---as defined under \texttt{symmetries}---to an irreducible representation associated with this group.
Depending on the group, the irreducible representation is expressed as
\begin{itemize}[label=\textbullet{}]
\item an integer/fraction (e.g.,~`\texttt{1}'/`\texttt{1/2}') denoting the helicity---Lorentz,
\item a partition (e.g.,~`\texttt{[2,1]}') denoting the Young Diagram---$SU(N)$,
\item an integer/fraction (e.g.,~`\texttt{1}'/`\texttt{1/2}') denoting the charge---$U(1)$.
\end{itemize}
The field is assumed to transform like a singlet under every group defined under \texttt{symmetries} that is not explicitly listed under \texttt{representations}.
Hence, every representation not explicitly defined assumes an appropriate default value (`\texttt{0}' for the helicity/charge and `\texttt{[]}' for the Young Diagram).

\item[{\parbox[b]{\linewidth}{%
\texttt{anticommute \Texinfocommandstyletextvar{Scalar}}}}]
\index[cp]{anticommute (field)@\texttt{anticommute \textnormal{(field)}}}%
If set to `\texttt{False}' (`\texttt{True}'), the field is treated as commuting (anticommuting). By default, this property is derived from the field's helicity, respecting spin-statistics.

\item[{\parbox[b]{\linewidth}{%
\texttt{conjugate \Texinfocommandstyletextvar{Scalar}}}}]
\index[cp]{conjugate (field)@\texttt{conjugate \textnormal{(field)}}}%
Whether to automatically include the conjugate field.
By default, this property is derived from the field's representations.
If there is at least one complex representation or an odd number of pseudo-real representations, the conjugate field is included as well.
Otherwise, only the field which is explicitly defined in the model file is included.

\item[{\parbox[b]{\linewidth}{%
\texttt{generations \Texinfocommandstyletextvar{Scalar}}}}]
\index[cp]{generations (field)@\texttt{generations \textnormal{(field)}}}%
The number of copies with the same representations appearing in the model.
The number of \texttt{generations} must be a positive integer.
By default, each field has just a single generation.

\item[{\parbox[b]{\linewidth}{%
\texttt{tex \Texinfocommandstyletextvar{Scalar}}\\
\texttt{tex\_hc \Texinfocommandstyletextvar{Scalar}}}}]
\index[cp]{tex (field)@\texttt{tex \textnormal{(field)}}}%
\index[cp]{tex\_hc (field)@\texttt{tex\_hc \textnormal{(field)}}}%
The \TeX{} string associated with the field and its conjugate.
By default, \texttt{tex} is set to the field name surrounded by \texttt{\textbackslash{}mathtt\{\textbullet{}\}}.
If \texttt{tex\_hc} is omitted, it is set to the value of \texttt{tex} and---depending on the value of \texttt{conjugate}---optionally appended by `\texttt{\^{}\textbackslash{}dagger}'.
\end{description}
\end{description}

\section{{Output Files}}
\label{anchor:Output}%

\begin{description}
\item[{\parbox[b]{\linewidth}{%
\texttt{version \Texinfocommandstyletextvar{Scalar}}}}]
\index[cp]{version (output)@\texttt{version \textnormal{(output)}}}%
The version of \texttt{autoeft} used to produce the file.

\item[{\parbox[b]{\linewidth}{%
\texttt{type \Texinfocommandstyletextvar{Sequence}}}}]
\index[cp]{type (output)@\texttt{type \textnormal{(output)}}}%
Details of the operator type.
The first entry denotes how often each field---identified by its name as defined under \texttt{fields} in the model file---and the covariant derivative appears in the type.
The second entry denotes if the type is `\texttt{real}' or `\texttt{complex}' (see \hyperref[anchor:Vocabulary]{[Vocabulary], page~\pageref*{anchor:Vocabulary}}).

\item[{\parbox[b]{\linewidth}{%
\texttt{generations \Texinfocommandstyletextvar{Mapping}}}}]
\index[cp]{generations (output)@\texttt{generations \textnormal{(output)}}}%
The number of generations for each field---identified by its name as defined under \texttt{fields} in the model file.

\item[{\parbox[b]{\linewidth}{%
\texttt{n\_terms \Texinfocommandstyletextvar{Scalar}}}}]
\index[cp]{n\_terms (output)@\texttt{n\_terms \textnormal{(output)}}}%
The total number of independent contractions with definite permutation symmetry of the repeated fields, not including multiple generations.

\item[{\parbox[b]{\linewidth}{%
\texttt{n\_operators \Texinfocommandstyletextvar{Scalar}}}}]
\index[cp]{n\_operators (output)@\texttt{n\_operators \textnormal{(output)}}}%
The total number of independent contractions with definite permutation symmetry of the repeated fields, including multiple generations.

\item[{\parbox[b]{\linewidth}{%
\texttt{invariants \Texinfocommandstyletextvar{Mapping}}}}]
\index[cp]{invariants (output)@\texttt{invariants \textnormal{(output)}}}%
Details of the type's invariant contractions.
Each entry is a mapping from a group name---as defined under \texttt{symmetries} in the model file---to the invariant contractions associated with this group.
Each independent contraction is labeled by `\texttt{O(<groupname>,<m>)}', where `\texttt{m}' enumerates the contractions.
The indices are denoted by `\texttt{<i>\_<j>}', where `\texttt{i}' is the position of the field that carries this index, and `\texttt{j}' is the position of the index \emph{on} the field.
Per invariant contraction, each index appears exactly twice and summation is implied.
For the Lorentz group, `\texttt{<i>\_<j>\~{}}' denotes dotted indices.
If the environment variable \texttt{AUTOEFT\_DS} is set, it replaces the tilde symbol `\texttt{\~{}}' in the above notation (see \hyperref[anchor:Environment]{[Environment], page~\pageref*{anchor:Environment}}).
Note that the dotted and undotted spinor indices of the building blocks (i.e.~fields plus derivatives) are understood to be (separately) symmetrized.
The symbol `\texttt{eps}' denotes the $\epsilon$-tensor with \texttt{eps(1,2)=eps(2\~{},1\~{})=1} for the Lorentz group and \texttt{eps(1,2,...,n)=1} for any internal $SU(N)$ group.
All indices not associated with the symmetry group in question are suppressed on the fields.
If the operators contain only fields that are singlets under a particular symmetry group, there is no index to be contracted and the entry contains just a single element `\texttt{+1}' multiplied by the fields without indices.

\item[{\parbox[b]{\linewidth}{%
\texttt{permutation\_symmetries \Texinfocommandstyletextvar{Sequence}}}}]
\index[cp]{permutation\_symmetries (output)@\texttt{permutation\_symmetries \textnormal{(output)}}}%
Details of the type's permutation symmetries.
\index[cp]{vector (permutation symmetry)@\texttt{vector \textnormal{(permutation symmetry)}}}%
The first entry is always a mapping from `\texttt{vector}' to a product of group names---as defined under \texttt{symmetries} in the model file---separated by `\texttt{\ * }'.
This denotes the order of the tensor product of the invariant contractions given under \texttt{invariants}.
All other entries represent explicit permutation symmetries.

\begin{description}
\item[{\parbox[b]{\linewidth}{%
\texttt{symmetry \Texinfocommandstyletextvar{Mapping}}}}]
\index[cp]{symmetry (permutation symmetry)@\texttt{symmetry \textnormal{(permutation symmetry)}}}%
The permutation symmetry of each field, identified by an integer partition (e.g.,~`\texttt{[2,1]}') corresponding to an irreducible representation of the symmetric group.

\item[{\parbox[b]{\linewidth}{%
\texttt{n\_terms \Texinfocommandstyletextvar{Scalar}}}}]
\index[cp]{n\_terms (permutation symmetry)@\texttt{n\_terms \textnormal{(permutation symmetry)}}}%
The number of independent contractions respecting the permutation symmetry, not including multiple generations.

\item[{\parbox[b]{\linewidth}{%
\texttt{n\_operators \Texinfocommandstyletextvar{Scalar}}}}]
\index[cp]{n\_operators (permutation symmetry)@\texttt{n\_operators \textnormal{(permutation symmetry)}}}%
The number of independent contractions respecting the permutation symmetry, including multiple generations.

\item[{\parbox[b]{\linewidth}{%
\texttt{matrix \Texinfocommandstyletextvar{Scalar}}}}]
\index[cp]{matrix (permutation symmetry)@\texttt{matrix \textnormal{(permutation symmetry)}}}%
The matrix representing the independent linear combinations of the invariant contractions respecting the permutation symmetry.
\end{description}
\end{description}

\section{{Vocabulary Glossary}}
\label{anchor:Vocabulary}%

\begin{description}
\item[{\parbox[b]{\linewidth}{%
\textsl{family}}}]
\index[cp]{family@family}%
\label{anchor:family}%
A family represents all operators with the same \textbf{Lorentz} representations of the fields (i.e.,~same \emph{kind} of fields).
For each field, the Lorentz representation can be identified with the helicity value $h$ and \texttt{autoeft} assigns the following symbols to each representation:

\begin{tabular}{m{.25\textwidth} m{.25\textwidth} m{.25\textwidth}}%
\toprule
object &helicity &symbol\\
\midrule
scalar               &$0$    &`\texttt{phi}'\\
spinor               &$-1/2$ &`\texttt{psiL}'\\
&$+1/2$ &`\texttt{psiR}'\\
rank-2 tensor        &$-1$   &`\texttt{FL}'\\
&$+1$   &`\texttt{FR}'\\
rank-4 tensor        &$-2$   &`\texttt{CL}'\\
&$+2$   &`\texttt{CR}'\\
covariant derivative &&`\texttt{D}'\\
\bottomrule
\end{tabular}%

Each family is then represented by a string consisting of these symbols, each preceded by its number of occurrences in the family and separated by an underscore `\texttt{\_}'.
To identify a family uniquely, the symbols are sorted by their helicity value and the covariant derivative is always added to the end.
Representations not appearing in the family are simply dropped.

For example, the family of the dimension-5 Weinberg operator---consisting of two spinors and two scalars---is given by `\texttt{2psiL\_2phi}' and its conjugate by `\texttt{2phi\_2psiR}'.

\item[{\parbox[b]{\linewidth}{%
\textsl{type}}}]
\index[cp]{type@type}%
\label{anchor:type}%
A type represents all operators with the same \textbf{Lorentz} and \textbf{internal} representations of the fields (i.e.,~same \emph{content} of fields).
Each type is then represented by a string consisting of the field name---as defined under \texttt{fields} in the model file---and each field is preceded by its number of occurrences in the type and separated by an underscore `\texttt{\_}'.
To identify a type uniquely, the fields are first sorted by their helicity value and fields with the same helicity are sorted by their name alpha-numerically.
The covariant derivative is always added to the end.
Representations not appearing in the type are simply dropped.

For example, the type of the dimension-5 Weinberg operator---consisting of two lepton doublets `\texttt{L}' and two Higgs doublets `\texttt{H}'---is given by `\texttt{2L\_2H}' and its conjugate by `\texttt{2H+\_2L+}'.

\item[{\parbox[b]{\linewidth}{%
\textsl{term}}}]
\index[cp]{term@term}%
\label{anchor:term}%
A term represents all operators with the same explicit contraction of the fields with the invariant tensor structures of the external and internal symmetries, retaining open generation indices.

For the number of terms to be unambiguous, a term is required to have a definite permutation symmetry for all repeated fields.
That means, that general terms are always decomposed into their irreducible representations under the symmetric group with respect to the repeated fields.

\item[{\parbox[b]{\linewidth}{%
\textsl{operator}}}]
\index[cp]{operator@operator}%
\label{anchor:operator}%
An operator is a particular instance of a \emph{term} with fixed generation indices for all fields.
Since any term has a definite permutation symmetry for any repeated field, the independent operators correspond to the independent components of the respective Wilson coefficient.
This means that the number of operators is equal to the independent degrees of freedom of the EFT (Effective Field Theory).

\item[{\parbox[b]{\linewidth}{%
\textsl{basis}}}]
\index[cp]{basis@basis}%
\label{anchor:basis}%
A valid operator \emph{basis} that can be processed further is represented by a directory containing the file \texttt{model.json} referencing the model, a (hidden) file \texttt{.autoeft} containing metadata, and the respective operator files (see \hyperref[anchor:Output]{[Output], page~\pageref*{anchor:Output}}).
Note that, while the files \texttt{model.json} and \texttt{.autoeft} must be contained in the top-level directory of the basis, the operator files can be structured in subdirectories.
By default, the directories called \texttt{basis} created by \texttt{autoeft construct} compose valid bases.

\item[{\parbox[b]{\linewidth}{%
\textsl{real \textnormal{family/type}}}}]
\index[cp]{real family/type@real \textnormal{family/type}}%
A \emph{real} family contains types that are either real or both the type and its conjugate are part of the same family.

A \emph{real} type contains terms that are either Hermitian or the Hermitian conjugate terms are not independent and can be expressed as a combination of the terms of the same type.

\item[{\parbox[b]{\linewidth}{%
\textsl{complex \textnormal{family/type}}}}]
\index[cp]{complex family/type@complex \textnormal{family/type}}%
A \emph{complex} family only contains complex types and the conjugate types are part of the distinct conjugate family.

A \emph{complex} type only contains terms that are not Hermitian and the Hermitian conjugate terms can be expressed as a combination of the terms of the distinct conjugate type.
\end{description}

\clearpage
\section*{{Index}}
\label{anchor:Index}%

\printindex[cp]

%- }}}
%- }}}
%- {{{ bibliography:

\bibliographystyle{utphys}
\bibliography{paper}

%- }}}

\end{document}